\documentclass[12pt,preprint]{aastex}

\usepackage{natbib}

\shorttitle{A ROBUST MEASURE OF TIDAL CIRCULARIZATION}
\shortauthors{Meibom \& Mathieu}

\begin{document}

\title{A ROBUST MEASURE OF TIDAL CIRCULARIZATION IN COEVAL BINARY POPULATIONS:\\
{THE SOLAR-TYPE SPECTROSCOPIC BINARY POPULATION IN THE OPEN CLUSTER M35
\altaffilmark{1}}}

\author{S{\o}ren Meibom\altaffilmark{2,4} and
Robert D. Mathieu\altaffilmark{3,4}}
\affil{Astronomy Department, University of Wisconsin - Madison,
    Madison, WI - 53706, USA}

\altaffiltext{1}{WIYN Open Cluster Study. XXII.}
\altaffiltext{2}{{\it meibom@astro.wisc.edu}}
\altaffiltext{3}{{\it mathieu@astro.wisc.edu}}
\altaffiltext{4}{Visiting Astronomer, Kitt Peak National Observatory,
National Optical Astronomy Observatory, which is operated by the
Association of Universities for Research in Astronomy, Inc. (AURA)
under cooperative agreement with the National Science Foundation.}


\begin{abstract} \label{abs}

We present a new homogeneous sample of 32 spectroscopic binary
orbits in the young ($\sim$ 150 Myr) main-sequence open cluster
M35. The distribution of orbital eccentricity vs. orbital period
($e-\log(P)$) displays a distinct transition from eccentric to
circular orbits at an orbital period of $\sim$ 10 days. The
transition is due to tidal circularization of the closest binaries.
The population of binary orbits in M35 provide a significantly
improved constraint on the rate of tidal circularization at an
age of 150 Myr. We propose a new and more robust diagnostic of
the degree of tidal circularization in a binary population based
on a functional fit to the $e-\log(P)$ distribution. We call this
new measure the {\it tidal circularization period}. The tidal
circularization period of a binary population represents the
orbital period at which a binary orbit with the most frequent
initial orbital eccentricity circularizes (defined as $e = 0.01$)
at the age of the population. We determine the tidal circularization
period for M35 as well as for 7 additional binary populations
spanning ages from the pre main-sequence ($\sim$ 3 Myr) to late
main-sequence ($\sim$ 10 Gyr), and use Monte Carlo error analysis
to determine the uncertainties on the derived circularization
periods. We conclude that current theories of tidal circularization
cannot account for the distribution of tidal circularization periods
with population age.

\end{abstract}



\section{INTRODUCTION}	\label{intro}

In a coeval population of binary stars the closest binaries
(separations $\lesssim 0.5$ AU) tend to have circular orbits,
while wider binaries have orbits with a distribution of non-zero
eccentricities. This trend holds true for binaries with early-type
\citep{nz03,pts98,mama92,gmm84} as well as late-type
\citep{mmd04,lst+02,mca+01,mdl+92} main-sequence components,
for binaries with evolved stellar components \citep{mm92},
and in the orbital eccentricity distribution of large planets
around solar-type stars (http://exoplanets.org). The transition
from eccentric to circular orbits is explained by dissipative
tidal interactions caused by the reaction of the binary components
to each other's gravitational field \citep[tidal
circularization:][]{hut81,zahn77,darwin1879}. The theory of
tidal interactions predicts that the timescale of circularization
depends strongly on stellar separation. Consequently, for a coeval
binary population, the transition from eccentric to circular orbits
should occur at a well-defined binary period. However, despite
the coeval nature of a star cluster, its population of short
period binaries may be small in number and consist of systems
with a distribution of initial eccentricities, masses, and
stellar angular momenta. Accordingly, the size and heterogeneity
of the binary population will add complexity and uncertainty
to the observational determination of the tidal circularization period.

Nonetheless, the transition between eccentric and circular
binary orbits provides an important constraint on the rate
of tidal circularization as a function of orbital period,
integrated over the lifetime of a binary population.
Furthermore, the distribution of tidal circularization periods with
population age enables us to study the evolution of tidal
circularization with time.

The success of theoretical models describing the efficiency
and the evolution of tidal circularization can be measured
by their ability to ``predict'' the tidal circularization periods for
binary populations of different ages. The differences between
present-day models lie in the mechanism by which energy and
angular momentum is transported between the binary components
and their orbits. Currently, three theoretical models exist
for the mechanism for tidal dissipation in late-type
main-sequence stars:
1) {\it The equilibrium tide theory} describes retardation of
the hydrostatic tidal bulge (equilibrium tide) due to the
coupling of the tidal flow to the motion of turbulent eddies
in the stellar convective envelope. This dissipative coupling
is assumed to be responsible for a phase shift between the
tidal bulge and the orbital motion of the binary stars.
Because of the phase shifts a tidal torque is
established between the two stars \citep{zahn77,hut81};
2) {\it The theory of dynamical tides} describes the excitation
and damping of gravity (g) waves in the radiative zones of stars
due to the tidal forcing by the companion star
\citep{sp83,zahn77,zahn75,zahn70}. This dissipation mechanism has been
successfully applied to explain circularization of early-type
main-sequence stars with radiative envelopes \citep{cc97,gmm84,zahn77},
and recently has also been applied to the radiative cores of
late-type stars \citep{sw02,ws02,gd98,tpn+98}.
Inclusion of resonance locking between the tidally forced
modes and stellar eigenmodes provides increased efficiency of the
tidal coupling \citep{sw02,ws02,ws01,ws99a,ws99b};
3) A pure {\it hydrodynamical mechanism} has been proposed by
\citet{tassoul00,tt88} and \citet{tt96}. Tassoul suggests
that hydrodynamical flows (large-scale meridional currents)
induced by lack of symmetry about the rotation axis in the
tidally perturbed star are responsible for the tidal torques
on the component stars. This mechanism has been controversial
on theoretical grounds \citep{rieutord92,rz97,tt97}.

Despite the developments in the theory of tidal circularization,
the models still cannot account for the observed periods of
circular orbits in solar-mass binary populations
\citep[e.g.][]{ws02,cc97,mdl+92}. To help along further progress
of theoretical modeling there is a need for high quality observational
data on large and coeval samples of binary stars, allowing for
accurate determination of the transition from eccentric to
circular orbits.

This paper presents a homogeneous ($M_{prim} \sim 1~M_{\odot}$)
sample of 32 spectroscopic binary orbits in the open cluster M35
(NGC 2168; $\alpha_{2000} = 6^{h}~9^{m}$, $\delta_{2000} =
24\degr~20\arcmin$; $l = 186.59\degr$, $b = 2.19\degr$, distance
$\simeq 850~pc$). With an age similar to the Pleiades but approximately
three times richer, M35 is a benchmark for stellar astrophysics
at $\sim$ 150 Myr (Deliyannis et al. 2004, in preparation). The
age makes M35 an important testing ground for binary stars that
have recently ended their pre main-sequence (PMS) phase where
they were larger and fully convective. Thus M35 sets the initial
state for main-sequence tidal circularization.

Section~\ref{obs} describes the observations, instrument, and
telescope used in the spectroscopic survey of M35. In Section~\ref{p-e}
we present the distribution of orbital eccentricity as a function
of orbital period in M35 (the $e-\log(P)$ diagram). In Section~\ref{tp}
we motivate and present a new robust method for determining the
period at which the most frequent binary orbits circularize.
We will call this the {\it tidal circularization period}. In Section
~\ref{comparing} we discuss and compare our new diagnostic to
the previous measure of tidal circularization, the tidal circularization
cutoff period. We compare the precision and accuracy performance
of both diagnostics and determine measurement uncertainties on
the tidal circularization period. In Section~\ref{tpm35} we
determine the tidal circularization period for the binary population
in M35, and in Section~\ref{7tp} we determine the circularization
period for a series of published populations of late-type unevolved
main-sequence binaries. In Section~\ref{evol} we carry out a
comparison between current theoretical predictions of the efficiency
and evolution of tidal circularization and the circularization
periods derived in Sections~\ref{tpm35} and ~\ref{7tp}.
Section~\ref{conclusions} summarizes and presents our conclusions.


\section{OBSERVATIONS} \label{obs}

M35 has been included in the WIYN Open Cluster Study (WOCS)
program since 1997. The radial-velocity survey of the cluster
will be described in detail in Meibom et al. (in preparation). We
summarize the relevant points here. The initial selection of $\sim
2000$ target stars was based on photometric membership in the
cluster and proper-motion membership studies to $V \lesssim 15$ by
\citet{ms86} and \citet{cudworth71}. Bjorkman \& Mathieu
(unpublished) completed astrometry and photometry to $V
\sim 17$. Our radial-velocity study includes stars from $V \simeq
13$ to $V \simeq 16.5$, corresponding to a range in stellar mass
from $\sim 1.4~M_{\odot}$ ($(B-V)_{0} \sim$ 0.37) to $\sim
0.7~M_{\odot}$ ($(B-V)_{0} \sim$ 1.1), with solar mass stars
at V $\sim$ 15 ((B-V) $\sim$ 0.58). Masses are derived using
the Yale (Y2) stellar evolution models \citep{ykd03}. The
cluster reddening ($E_{(B-V)} = 0.20$) was adopted from Deliyannis
et al. 2004 (in preparation).

All spectroscopic data were obtained using the WIYN\footnote{
The WIYN Observatory is a joint facility of the University of
Wisconsin-Madison, Indiana University, Yale University, and the
National Optical Astronomy Observatories.} 3.5m telescope at
Kitt Peak, Arizona, USA. The telescope is equipped with a
Multi-Object Spectrograph (MOS) consisting of a fiber optic
positioner (Hydra) feeding a bench mounted spectrograph. The
Hydra positioner is capable of placing $\sim$ 95 fibers in a
1-degree diameter field with a precision of $0.2\arcsec$. In
the field of M35 approximately 82-85 fibers are positioned on
stars while the remaining fibers are used for measurements of
the sky background. We use the $3\arcsec$ diameter fibers
optimized for blue transmission, and the spectrograph is
configured with an echelle grating and an all-transmission
optics camera providing high throughput at a resolution of
$\sim$ 20,000. All observations were done at central wavelengths
of $5130\AA$ or $6385\AA$ with a wavelength range of $\sim 200\AA$,
providing rich arrays of narrow absorption lines. Radial velocities
with a precision of $\sim 0.5~km s^{-1}$ \citep{mbd+01} are derived
from the spectra via cross-correlation with a high $S/N$ sky spectrum.

Telescope time granted from NOAO\footnote{NOAO is the national
center for ground-based nighttime astronomy in the United States
and is operated by the Association of Universities for Research
in Astronomy (AURA), Inc. under cooperative agreement with the
National Science Foundation.} and Wisconsin allowed for 3-4 observing
runs per year for M35, with each run typically including
multiple observations on several sequential nights. Once identified,
velocity variables are observed at a frequency appropriate to the
timescale of their variation. At present the radial-velocity survey
has resulted in a sample of 81 spectroscopic binaries: 32 members,
37 candidate members and 12 non-members. Of the 32 member binaries,
25 are single-lined (SB1) and 7 are double-lined (SB2) spectroscopic
binaries. The orbital periods span from 2.25 days to 1115 days
among the member binaries, corresponding to separations from
$\sim$ 0.04 to 2.5 AU assuming a $1~M_{\odot}$ primary and a
$0.5~M_{\odot}$ secondary component. The 32 member binaries
were identified in the 2MASS All Sky Survey based on equatorial
coordinates. 2MASS ID's, astrometry, photometry, and orbital
parameters are listed in Table 1.

\begin{deluxetable}{cccccccccccccccc}
\tabletypesize{\scriptsize}
\rotate
\tablecaption{Astrometric, photometric and orbital data for
binary members of M35. \label{tbl-1}}
\tablewidth{0pt}
\tablehead{
\colhead{2MASS} & \colhead{MS} & \colhead{Cd} & \colhead{$\alpha$ (2000)} & \colhead{$\delta$ (2000)} &
\colhead{$V$} & \colhead{$B-V$} & \colhead{Binary} & \colhead{\# RV's} &
\colhead{$\gamma$} & \colhead{Period} & \colhead{$e$} &
\colhead{MP} & \colhead{$M_{prim}$} &
\colhead{$M_{sec}$} & \colhead{Spectral}\\
\colhead{ID\tablenotemark{1}} & \colhead{ID\tablenotemark{2}} & \colhead{ID\tablenotemark{3}} & \colhead{h m s} & \colhead{$\degr$ $\arcmin$ $\arcsec$} &
\colhead{} & \colhead{} & \colhead{type} & \colhead{} &
\colhead{($km s^{-1}$)} & \colhead{(days)} & \colhead{} &
\colhead{(\%)\tablenotemark{4}} & \colhead{($M_{\odot}$)} &
\colhead{($M_{\odot}$)} & \colhead{type}}
\startdata
06090257+2420447 & 249 & 402 & 06 09 02.550 & +24 20 44.62  & 13.62 & 0.60 & SB1 &    32 & $-$8.3  &   10.280 &  0.009 $\pm$0.019 &  99 & 1.32 & ...  & F7 \\
06090403+2423483 & 252 & 404 & 06 09 04.010 & +24 23 48.26  & 13.31 & 0.63 & SB2 &  11/9 & $-$8.7  &   22.490 &  0.155 $\pm$0.017 &  99 & 1.28 & 1.01 & M  \\
06090521+2419309 & 259 & 410 & 06 09 05.180 & +24 19 30.92  & 13.32 & 0.60 & SB1 &    25 & $-$8.4  &    2.703 &  0.028 $\pm$0.028 &  99 & 1.32 & ...  & F7 \\
06090650+2413499 & 267 & 419 & 06 09 06.470 & +24 13 49.90  & 14.62 & 0.75 & SB1 &    18 & $-$6.8  &  344.6   &  0.334 $\pm$0.030 &  97 & 1.05 & ...  & G0 \\
06091099+2421515 & 301 & 447 & 06 09 10.960 & +24 21 51.51  & 13.65 & 0.62 & SB1 &    20 & $-$7.7  &   58.02  &  0.502 $\pm$0.029 &  97 & 1.28 & ...  & F8 \\
06092037+2412177 & 360 & 512 & 06 09 20.340 & +24 12 17.74  & 13.64 & 0.62 & SB2 &  15/7 & $-$8.5  &   35.380 &  0.272 $\pm$0.017 &  99 & 1.28 & 0.83 & M  \\
06093861+2417394 & 463 & 612 & 06 09 38.590 & +24 17 39.54  & 14.38 & 0.73 & SB1 &    18 & $-$8.1  &   79.28  &  0.375 $\pm$0.010 &  94 & 1.05 & ...  & G0 \\
06083426+2421359 & ... & ... & 06 08 34.220 & +24 21 35.73  & 16.03 & 1.16 & SB2 &  18/7 & $-$7.9  &   56.14  &  0.353 $\pm$0.026 & ... & 0.77 & 0.77 & M  \\
06085047+2419382 & ... & ... & 06 08 50.440 & +24 19 38.16  & 16.42 & 1.26 & SB2 & 19/14 & $-$5.3  &   24.108 &  0.219 $\pm$0.013 & ... & 0.69 & 0.66 & M  \\
06090352+2417234 & ... & ... & 06 09 03.490 & +24 17 23.36  & 15.95 & 1.10 & SB1 &    18 & $-$7.5  &  156.9   &  0.594 $\pm$0.024 & ... & 0.77 & ...  & K1 \\
06091557+2410422 & ... & ... & 06 09 15.540 & +24 10 42.16  & 15.47 & 1.03 & SB1 &    36 & $-$7.9  &    8.170 &  0.538 $\pm$0.013 & ... & 0.77 & ...  & K1 \\
06091924+2417223 & ... & ... & 06 09 19.210 & +24 17 22.37  & 15.53 & 0.92 & SB1 &    18 & $-$8.8  &  795     &  0.208 $\pm$0.045 & ... & 0.86 & ...  & G7 \\
06092436+2426200 & ... & ... & 06 09 24.330 & +24 26 20.03  & 15.34 & 0.88 & SB1 &    22 & $-$7.4  &   10.330 &  0.016 $\pm$0.009 & ... & 0.95 & ...  & G3 \\
06084130+2426389 & ... & ... & 06 08 41.260 & +24 26 38.88  & 16.10 & 1.14 & SB1 &    18 & $-$6.9  &   18.426 &  0.276 $\pm$0.016 & ... & 0.77 & ...  & K1 \\
06090444+2412441 & ... & ... & 06 09 04.410 & +24 12 44.00  & 14.86 & 0.80 & SB1 &    15 & $-$9.1  &   49.075 &  0.358 $\pm$0.018 & ... & 0.95 & ...  & G3 \\
06093267+2415041 & ... & ... & 06 09 32.640 & +24 15 04.14  & 14.26 & 0.67 & SB1 &    12 & $-$7.46 &    7.088 &  0.003 $\pm$0.005 & ... & 1.19 & ...  & F9 \\
06104368+2416089 & ... & ... & 06 10 43.680 & +24 16 09.05  & 14.50 & 0.82 & SB1 &    17 & $-$9.5  &   10.077 &  0.008 $\pm$0.005 &  72 & 0.95 & ...  & G3 \\
06101134+2426415 & ... & ... & 06 10 11.350 & +24 26 41.42  & 15.32 & 1.02 & SB1 &    32 & $-$6.4  &   14.524 &  0.246 $\pm$0.044 & ... & 0.86 & ...  & G7 \\
06095563+2417454 & ... & ... & 06 09 55.627 & +24 17 45.60  & 15.02 & 0.88 & SB1 &    18 & $-$7.5  &   30.130 &  0.273 $\pm$0.005 & ... & 0.95 & ...  & G3 \\
06094745+2423085 & 494 & 650 & 06 09 47.455 & +24 23 08.56  & 13.63 & 0.67 & SB1 &    15 & $-$6.9  & 1114.97  &  0.543 $\pm$0.072 &  90 & 1.19 & ...  & F9 \\
06092221+2446528 & ... & ... & 06 09 22.217 & +24 46 52.68  & 15.00 & 0.78 & SB1 &    15 & $-$8.6  &    8.010 &  0.041 $\pm$0.032 &  72 & 0.95 & ...  & G3 \\
06092708+2413452 & 404 & 550 & 06 09 27.094 & +24 13 45.12  & 13.80 & 0.73 & SB2 & 20/15 & $-$10.2 &    4.442 &  0.010 $\pm$0.006 &  98 & 1.05 & 0.92 & M  \\
06092536+2404037 & ... & 541 & 06 09 25.363 & +24 04 03.72  & 13.76 & 0.61 & SB2 & 24/20 & $-$7.5  &   16.488 &  0.041 $\pm$0.014 &  57 & 1.28 & 1.13 & M  \\
06085334+2432309 & 198 & 343 & 06 08 53.335 & +24 32 30.80  & 14.32 & 0.67 & SB1 &    15 & $-$8.0  &   22.619 &  0.404 $\pm$0.007 &  96 & 1.19 & ...  & F9 \\
06085441+2403081 & ... & ... & 06 08 54.418 & +24 03 08.06  & 14.97 & 0.77 & SB1 &    19 & $-$7.45 &   12.280 &  0.550 $\pm$0.003 & ... & 1.05 & ...  & G0 \\
06083296+2408164 & ... & ... & 06 08 32.969 & +24 08 16.30  & 15.05 & 0.78 & SB1 &    17 & $-$7.3  &  457.0   &  0.512 $\pm$0.064 & ... & 0.95 & ...  & G3 \\
06083082+2417547 & ... & ... & 06 08 30.818 & +24 17 54.67  & 15.47 & 0.88 & SB2 & 16/15 & $-$10.3 &   18.590 &  0.429 $\pm$0.009 & ... & 0.95 & 0.89 & M  \\
06083184+2410384 &  75 & 229 & 06 08 31.841 & +24 10 38.35  & 14.57 & 0.74 & SB1 &    18 & $-$7.1  &  475     &  0.507 $\pm$0.021 &  89 & 1.05 & ...  & G0 \\
06080751+2423413 &   9 & 122 & 06 08 07.500 & +24 23 41.46  & 13.94 & 0.60 & SB1 &    19 & $-$8.3  &  930     &  0.474 $\pm$0.079 &  98 & 1.32 & ...  & F7 \\
06074436+2430262 & ... & ... & 06 07 44.352 & +24 30 26.10  & 15.23 & 0.82 & SB1 &    11 & $-$6.9  &  473     &  0.309 $\pm$0.121 & ... & 0.95 & ...  & G3 \\
06100456+2437000 & ... & ... & 06 10 04.562 & +24 37 00.19  & 15.74 & 1.03 & SB1 &    16 & $-$8.3  &   12.566 &  0.095 $\pm$0.006 & ... & 0.86 & ...  & G7 \\
06083789+2431455 & ... & ... & 06 08 37.889 & +24 31 45.48  & 15.95 & 1.05 & SB1 &    14 & $-$8.1  &    2.247 &  0.010 $\pm$0.008 & ... & 0.77 & ...  & K1 \\
\enddata

\tablerefs{(1) 2MASS ID's; (2) \citet{ms86} ID's; (3) \citet{cudworth71} ID's; (4) Proper motion membership probability by \citet{ms86,cudworth71}}

\end{deluxetable}


\section{THE PERIOD-ECCENTRICITY DISTRIBUTION IN M35} \label{p-e}

\begin{figure}
\epsscale{1.0}
\plotone{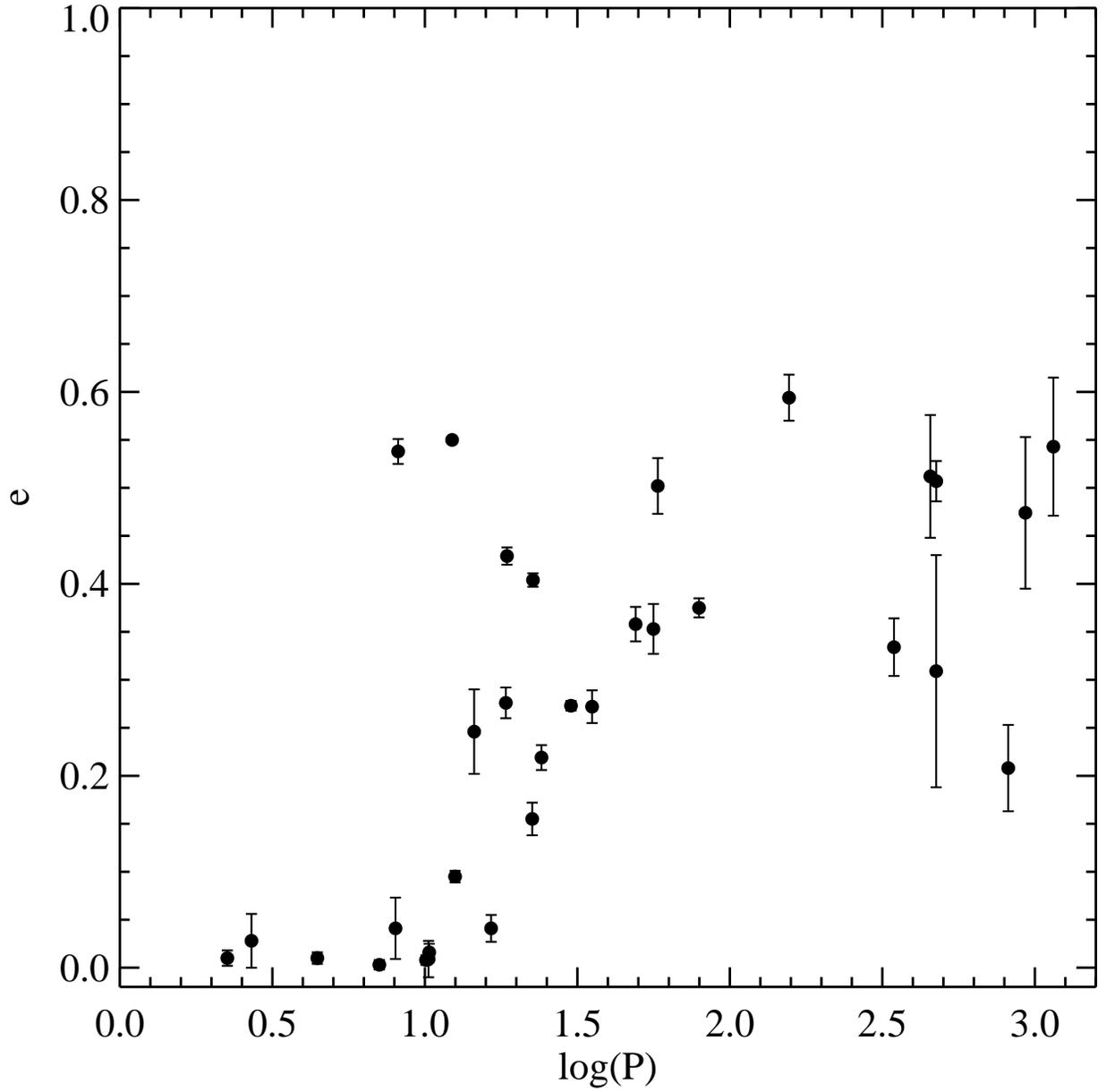}
\caption{The distribution of orbital eccentricity (e) as a
function of $log$ orbital period (P) for 32 solar-type
spectroscopic binary members in M35.
\label{m35_log}}
\end{figure}

The new compilation of binary orbits in M35 greatly improves
the constraint on tidal circularization at the early stage
of the main-sequence phase. Prior to this publication, the
small sample of binary orbits in the Pleiades provided only
rough limits on the transition between eccentric and circular 
orbits at this age \citep{mrd+92}. Figure~\ref{m35_log}
shows the distribution of orbital eccentricity ($e$) as a
function of $log$ orbital period (P) for the 32 spectroscopic
binary members of M35. The shortest period binaries have
orbital eccentricities consistent with circular orbits,
while the orbits for longer period binaries show a distribution
of non-zero eccentricities. The presumably primordial
distribution of eccentricities for the longest period
binaries is consistent with the Gaussian distribution observed
in late-type main-sequence binary populations in other open
clusters \citep{dmm92}, in the solar neighborhood \citep{dm91},
and from the Galactic halo \citep{lst+02}. The lack of high
eccentricity orbits ($e \gtrsim 0.6$) among the long period
binaries in M35 may be due in part to observational biases.

\begin{figure}
\epsscale{1.0}
\plotone{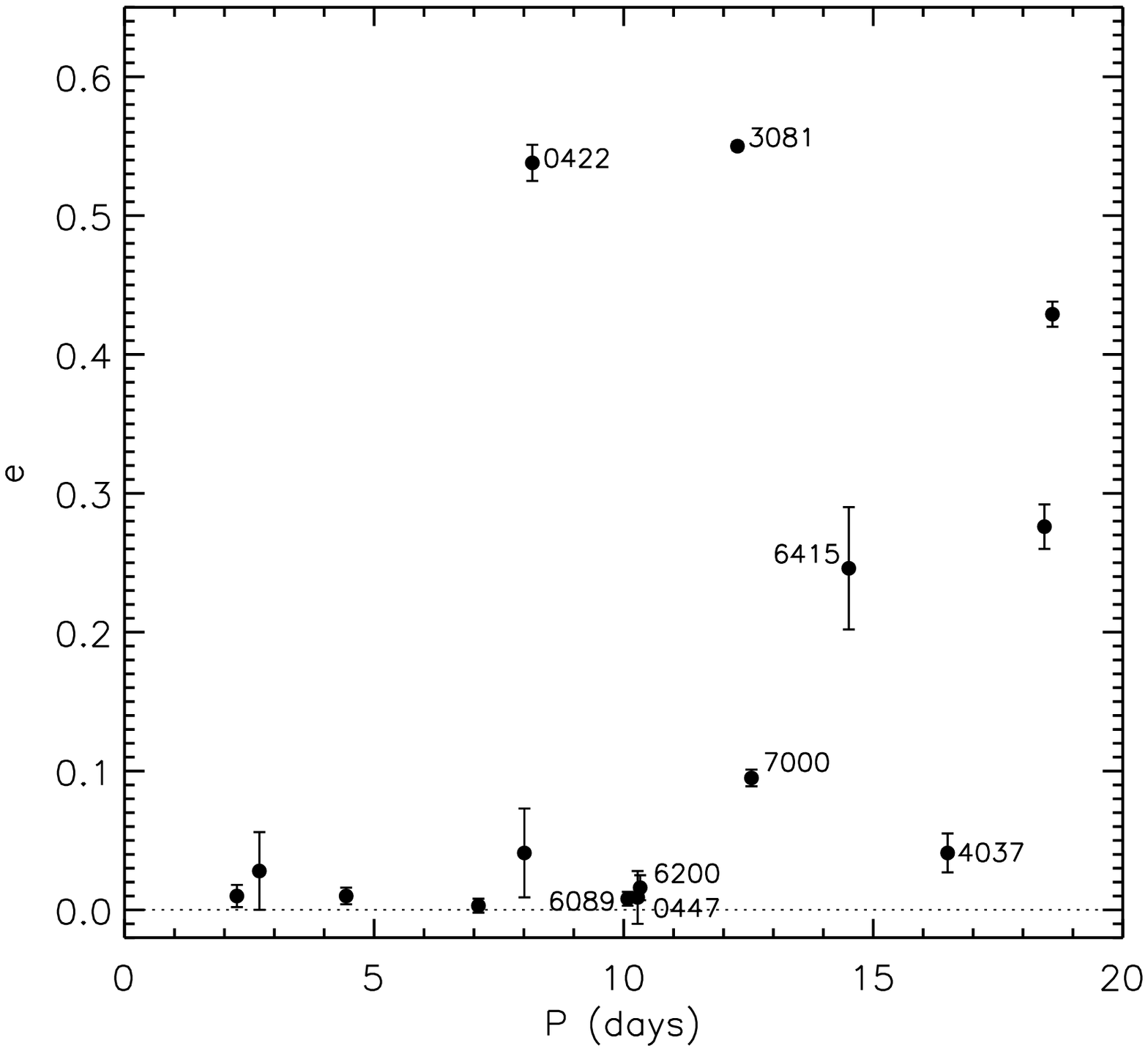}
\caption{The eccentricity distribution of M35 binary orbits
with periods shorter than 20 days. Numbers next to selected
data refer to binary ID numbers in Table 1.
\label{m35_lin_z}}
\end{figure}

All binaries whose periods and eccentricities are plotted
in Figure~\ref{m35_log} are cluster members based on their
radial velocities and location in the color-magnitude diagram
(CMD). Fifteen have confirmation of membership from proper-motion
measurements. A few binary systems deserve special attention
and Figure~\ref{m35_lin_z} shows on a linear scale the distribution
of orbital eccentricity vs. period for binaries with periods less
than 20 days. The last 4 digits of the 2MASS ID are used to
label key binaries in Figure~\ref{m35_lin_z} and for reference
in the text and in Table 1. Nine binary orbits have eccentricities
less than $3 \times \sigma_{e}$ above $e = 0.0$, where $\sigma_{e}$
denotes the error on the orbital eccentricity. We consider these
binary orbits to be circular. Binary 4037 ($P_{orb} = 16.49$ days,
$e = 0.041 \pm 0.014$) has the longest period among the binaries
with circular orbits. Three circular binaries have orbital periods
close to 10 days: 6089 ($P_{orb} = 10.08$ days, $e = 0.008 \pm
0.005$), 0447 ($P_{orb} = 10.28$ days, $e = 0.009 \pm 0.019$),
and 6200 ($P_{orb} = 10.33$ days, $e = 0.016 \pm 0.009$).
4 binaries have eccentric orbits with periods shortward of
the longest period circular orbit: 0422 ($P = 8.17$ days,
$e = 0.538 \pm 0.013$), 3081 ($P = 12.28$ days, $e = 0.550
\pm 0.003$), 7000 ($P = 12.57$ days, $e = 0.095 \pm 0.006$),
and 6415 ($P = 14.52$ days, $e = 0.246 \pm 0.044$).

Thus regardless of the homogeneity in age, metallicity and primary
mass of the M35 binary population there is a $\sim$ 8 day overlap
in period between eccentric and circular binary orbits.


\section{DETERMINING THE CIRCULARIZATION PERIOD} \label{tp}

The transition from eccentric to circular orbits observed
in M35 is also found in observational studies of late-type
main-sequence binary populations in well known open clusters
such as the Pleiades \citep{mrd+92}, the Hyades/Praesepe
\citep{dmm92,mm99}, M67 \citep{lmm92}, and NGC188 \citep{mmd04},
and in populations of PMS \citep[and references therein]{mca+01},
field \citep{dm91}, and halo \citep{lst+02} binaries.
This characteristic fingerprint of tidal circularization
is an important observational constraint on the theory of
tidal circularization. Consequently, determination of the
period at which binary orbits becomes circular in a given
binary population is critical to constraining theoretical
models. However, determining this tidal circularization period
is not straightforward in any of the above populations.

The longest period circular orbit has been the preferred measure
over the past decade \citep[e.g.][]{dmm92} and has been referred
to as the {\it tidal circularization cutoff period} (hereinafter:
cutoff period). We argue here that the cutoff period is not the
optimal measure of the transition between eccentric and circular
orbits.

Populations of short period binaries vary in size and consist
of systems with a distribution of initial eccentricities.
This is observed in the Gaussian distribution of orbital
eccentricity at longer orbital periods where presumably orbital
evolution has been minimal \citep{hmu+03,dmm92,dm91}.
The Gaussian distribution of initial orbital eccentricities
combined with the effects of tidal circularization will lead
to an overlap in period-space between initially low eccentricity
orbits that have been circularized and initially high eccentricity
orbits not yet circularized. This overlap is indeed observed in all
published populations of coeval binary orbits with the exception of
the small sample in the Pleiades cluster (see Section~\ref{7tp},
Figure~\ref{all_log}). The width of this overlap was estimated by
\citet{mm88} based on the observed distribution of initial orbital
eccentricities, and indeed, \citet{dmm92} showed that the existence
of eccentric orbits in the Hyades/Praesepe populations with periods
shorter than the longest period circular orbit can be explained
due to incomplete tidal circularization of binaries with initially
high orbital eccentricity. We similarly find through simulations
of tidal circularization using the equilibrium tide theory by
\citet{zahn77} that eccentric orbits with periods shorter than
the longest period circular orbit can be a result of high initial
eccentricity ($e \gtrsim 0.5$).

For short-period binaries with low but non-zero orbital eccentricity
($0.0 \lesssim e \lesssim 0.1$), \citet{mazeh90} suggests that the
eccentricity is induced by the presence of a third companion star
due to dynamical interactions different from tidal effects. However,
most eccentric orbits with periods shorter than the longest
period circular orbit have eccentricities between 0.1 and 0.6,
and among the 4 binaries in M35 with moderate to significant
eccentricities and periods shortward of the longest period circular
orbit we see no indications in the stellar spectra of a third component.

\begin{figure}
\epsscale{1.0}
\plotone{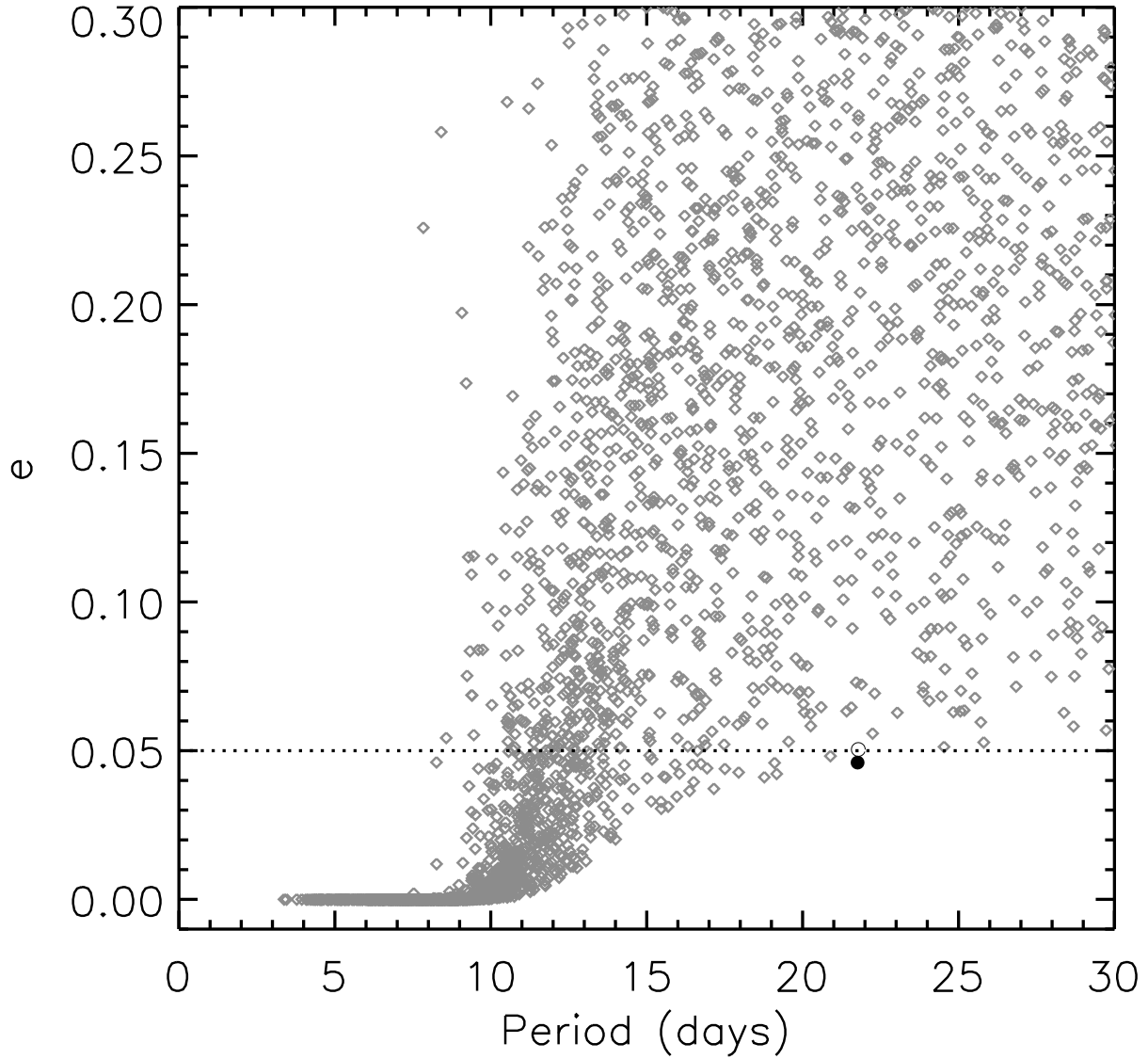}
\caption{The $e-\log(P)$ diagram for a tidally circularized population
of artificially generated binaries (grey diamonds). The lower limit
on the initial orbital eccentricity is shown as a dotted line at
$e = 0.05$. The initial and final locations of the longest period
circular orbit are marked with an open and a filled circle, respectively.
\label{piei_pfef}}
\end{figure}

Monte Carlo simulations of tidal circularization using the equilibrium
tide theory by \citet{zahn77} show that the longest period circular
binary orbit defining the cutoff period is likely to originate
from the low eccentricity tail of the initial Gaussian eccentricity
distribution. Accordingly, the cutoff period does not measure tidal
circularization of the most frequent binary orbit. We show in
Figure~\ref{piei_pfef} the $e-\log(P)$ diagram of a tidally
circularized population of artificially generated binaries (grey
diamonds). To focus on the process of tidal circularization, the
lower limit on the initial orbital eccentricity was set to 0.05
(dotted horizontal line). The threshold of 0.05 is 2-3 times
the typical error on orbital eccentricities determined from
observations. The initial and final locations of the longest
period circular orbit are marked with an open and a filled circle,
respectively.

Figure~\ref{piei_pfef} also demonstrate that the probability
of measuring the ``true'' cutoff period of a parent population
of binaries is highly sensitive to the size of the observed
binary population. Consequently, the value of the observed cutoff
period will vary with the size of the population.

Furthermore, long period (P $\sim 2 \times 10^1 - 10^2$ days)
circular ($e < 0.05$) or marginally circular ($e \lesssim 0.1$)
orbits of unevolved binaries are observed in the majority of the
published binary populations (see Section~\ref{7tp}, Figure~\ref{all_log}),
indicating the possibility that binary stars can also form with
very low eccentricity. While not included in Figure~\ref{piei_pfef},
the cutoff period is vulnerable to such primordial circular orbits.

Finally, the determination of the cutoff period makes no use of
the information provided by eccentric orbits in the $e-\log(P)$ diagram.

Consequently we suggest that the cutoff period is not a robust measure of
the state of tidal circularization in a coeval binary population.

So motivated, we propose a new method for determining the
period of circularization of primordial eccentric binary
orbits. We refer to this period as the {\it tidal circularization
period}. Our proposed method will make use of the information
provided by all binary orbits in a population by fitting to
the observed period-eccentricity distribution a function of
the form:

\begin{equation}
e(P) = \left\{ \begin{array}{ll}
0.0 & \textrm{if $P \le P\arcmin$}\\
\alpha \bigl(1 - e^{\beta(P\arcmin - P)}~\bigr)^\gamma & \textrm{if $P > P\arcmin$}
\end{array} \right.
\end{equation}

\begin{figure}
\epsscale{1.0}
\plotone{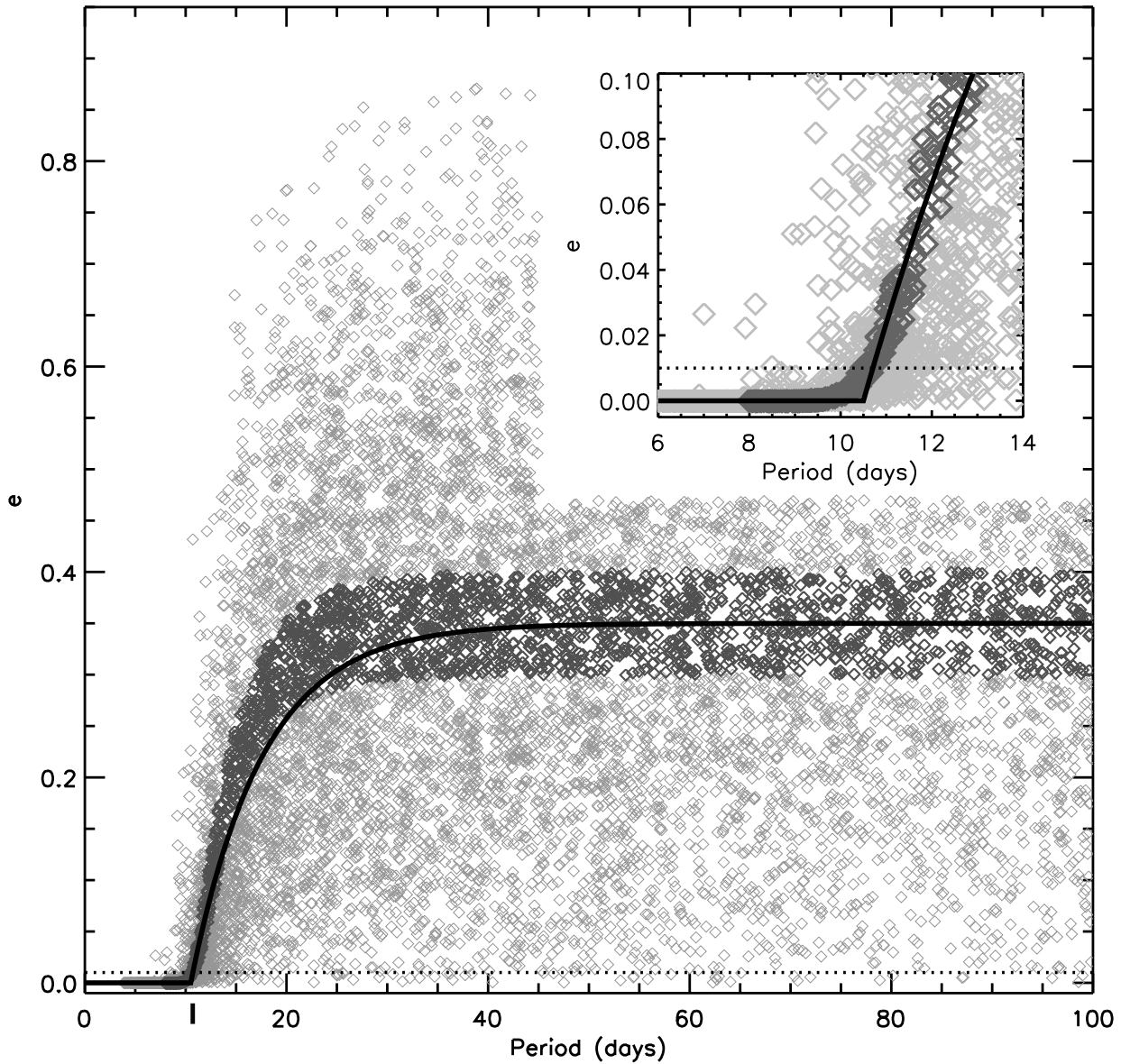}
\caption{The circularization function (black solid line) fitted
to an artificially generated sample of binary orbits (diamonds).
Tidal circularization of these binary orbits has been simulated
by numerical integration of the differential equations (3) and (4).
Selection of the initial orbital periods and the initial orbital
eccentricities is described in the main text. Binaries with the
``typical'' (most frequent) initial eccentricities between 0.3
and 0.4 are highlighted (dark diamonds). The circularization
period (CP) is defined as the period at which the circularization
function has a value of 0.01 ($e(CP) = 0.01$). $e = 0.01$ is marked
by a horizontal dotted line and CP is marked by the vertical
black line on the period axis. The insert is a close-up view of
the transition-region.
\label{parent_new_cp=10d}}
\end{figure}

This mathematical function is motivated by the observed 
period-eccentricity distributions and by numerical modeling
using the theory of Zahn (1977). The function is not physically
derived but constructed to mimic the tidal circularization
isochrone of the most frequently occurring eccentric binary
orbits. Figure~\ref{parent_new_cp=10d} shows the function
in the $e-\log(P)$ diagram.  The transition of the function
from eccentric to circular orbits is managed by the
$e^{\beta(P\arcmin - P)}$ term and the $\gamma$ coefficient.
$\gamma$ controls the abruptness of the break from $e=0$
at the period $P\arcmin$ and $\beta$ controls the overall
steepness/slope of the transition. Our choices of $\beta$
and $\gamma$ are discussed in Section~\ref{comparing}. The
value of $\alpha$ ($\alpha = 0.35$) is set to ensure that
$e(P > P\arcmin)$ approaches a value of 0.35, the mean eccentricity
of all observed binary orbits with periods longer than 50 days
in the Pleiades, M35, Hyades/Praesepe, M67, and NGC188. For
periods shortward of $ P\arcmin$ the function is set to
0.0 ($e(P < P\arcmin) = 0.0$). We will hereafter refer to
this function as the circularization function. 

The value of $P\arcmin$ determines the location of the
onset of the rise in the circularization function. We
determine the location of the circularization function
in period by minimizing the total absolute deviation
($\delta$) between the observed ($e_i$) and the calculated
($e(P_i)$) eccentricities,

\begin{equation}
\delta = \sum_{i=1}^{N} \mid e_i - e(P_i) \mid ^{\eta}.
\end{equation}

\noindent $N$ denotes the number of binary orbits.
The value of the exponent $\eta$ controls the influence
on $P\arcmin$ by the binary orbits with maximum deviation from
the circularization function. We found that $\eta > 1$ causes
high sensitivity to the short period eccentric binaries, often
leading to values of $P\arcmin$ smaller than the period of
the shortest period eccentric orbit. $\eta = 1$ leads to
values of $P\arcmin$ between the periods of the shortest
period eccentric orbit and the longest period circular orbit.
We use $\eta = 1$ when fitting the circularization function
throughout this paper.

Figure~\ref{parent_new_cp=10d} shows the circularization function
fitted to an artificially generated sample of binary orbits. Tidal
circularization of these binary orbits has been simulated by numerical
integration of differential equations derived using equations
(4.3) and (4.4) in \citet{zahn77} (Note erratum by \citet{zahn78})
and the assumptions by \citet{dmm92}:

\begin{equation}
\frac{de}{dt} = \frac{-e}{A P^{16/3}}
\end{equation}

\begin{equation}
\frac{dP}{dt} = \frac{-3 e^2}{A P^{13/3}}
\end{equation}

The constant $A$ controls the effectiveness of tidal dissipation
and thus the rate of tidal circularization. We set $A$ so that
numerical simulation of tidal circularization using equations
(3) and (4) produce transitions from eccentric to circular orbits
at periods similar to those observed.

The circularization function mimics the transition from
eccentric to circular orbits of the ``typical''binary ($0.3 <
e_{ini} < 0.4$; highlighted in Figure~\ref{parent_new_cp=10d}).
The tidal circularization isochrone for the typical binary
is reproduced by the circularization function at $e = 0.01$.
Thus we define the ``tidal circularization period''
(hereinafter circularization period or CP) as the period
for which the best fit circularization function has a value
of 0.01 ($e(CP) = 0.01$). The choice of $e=0.01$ as the
threshold for circularization is also similar to the
threshold chosen in theoretical simulations (see
Section~\ref{evol}), and thus facilitates direct comparison
between observations and theory.

Our estimate of the initial eccentricity of a binary orbit that
circularizes with a period equal to the circularization period
at the age of a binary population, is tied to the tidal theory
of \citet{zahn77}. Deviations from the typical (most frequent)
initial eccentricity depends on the value of the circularization
period (the age of the binary population). By fitting the
circularization function to rich distributions of tidally
circularized binary orbits (as shown in Figure~\ref{parent_new_cp=10d}),
we find that over the range of circularization periods of
interest to this study, $\sim$ 5 days to 15 days, the range
of initial eccentricities of the binaries represented by the
resulting circularization periods is 0.2-0.5. This range of
initial orbital eccentricities is centered on the mean of the
observed Gaussian eccentricity distribution ($\bar{e} = 0.35$)
and we conclude that according to the \citet{zahn77} theory
the circularization period represents the orbital period at
which a binary orbit with the most frequent initial orbital
eccentricity circularizes. The range of initial eccentricities
from 0.2 to 0.5 also includes the initial eccentricity used
in theoretical predictions of tidal circularization (see
Section~\ref{evol}) facilitating direct comparison between
these predictions and the observed circularization periods.

\section{EVALUATION OF THE PERFORMANCE OF THE TIDAL CIRCULARIZATION PERIOD} \label{comparing}

In the following we test the performance of the circularization
function and compare its performance to the cutoff period. Using a Monte
Carlo approach, we generate 10,000 binary populations each with
20 binaries with periods between 10 and 50 days, corresponding
to the number of known binary orbits in the M35 sample with orbital
periods shorter than 50 days. Tidal circularization was simulated
by numerical integration of the \citet{zahn77} differential equations
(eqs. [3] and [4] above). Initial orbital periods and eccentricities
for each population were determined by random selection from a
log-normal period distribution and a normal (Gaussian) eccentricity
distribution. \footnote{Orbits randomly selected with an initially
negative eccentricity were excluded.} We adopted the period distribution
of \citet{dm91} derived from solar-type binaries in the solar neighborhood,
while the Gaussian eccentricity distribution ($\bar{e} = 0.35, \sigma =
0.21$) was determined by fitting the distributions of all orbital
eccentricities from the Pleiades, M35, Hyades/Praesepe, M67, and
NGC188 for orbits with periods longer than 50 days.

We ran the Monte Carlo experiment twice. The first time we
excluded initially circular orbits ($e_{i} < 0.05$) to test
the performance of the diagnostics in the absence of initially
circular binaries. The second time we allowed for initially
circular orbits to see the effect of such systems on the
circularization and cutoff periods.

The distributions of circularization periods and cutoff periods
resulting from the two Monte Carlo experiments described above
allow us to determine the uncertainty on each of these measures.
We will refer to the circularization period and the cutoff period
derived from each binary population as the {\it observed}
circularization period and the {\it observed} cutoff period.
For reference two ``parent'' populations of $\sim 20,000$ binary
orbits were generated and tidally evolved. We will refer to
the circularization period and the cutoff period derived
from these parent populations as the {\it true} circularization
period and the {\it true} cutoff period.

\begin{figure}
\epsscale{0.75}
\plotone{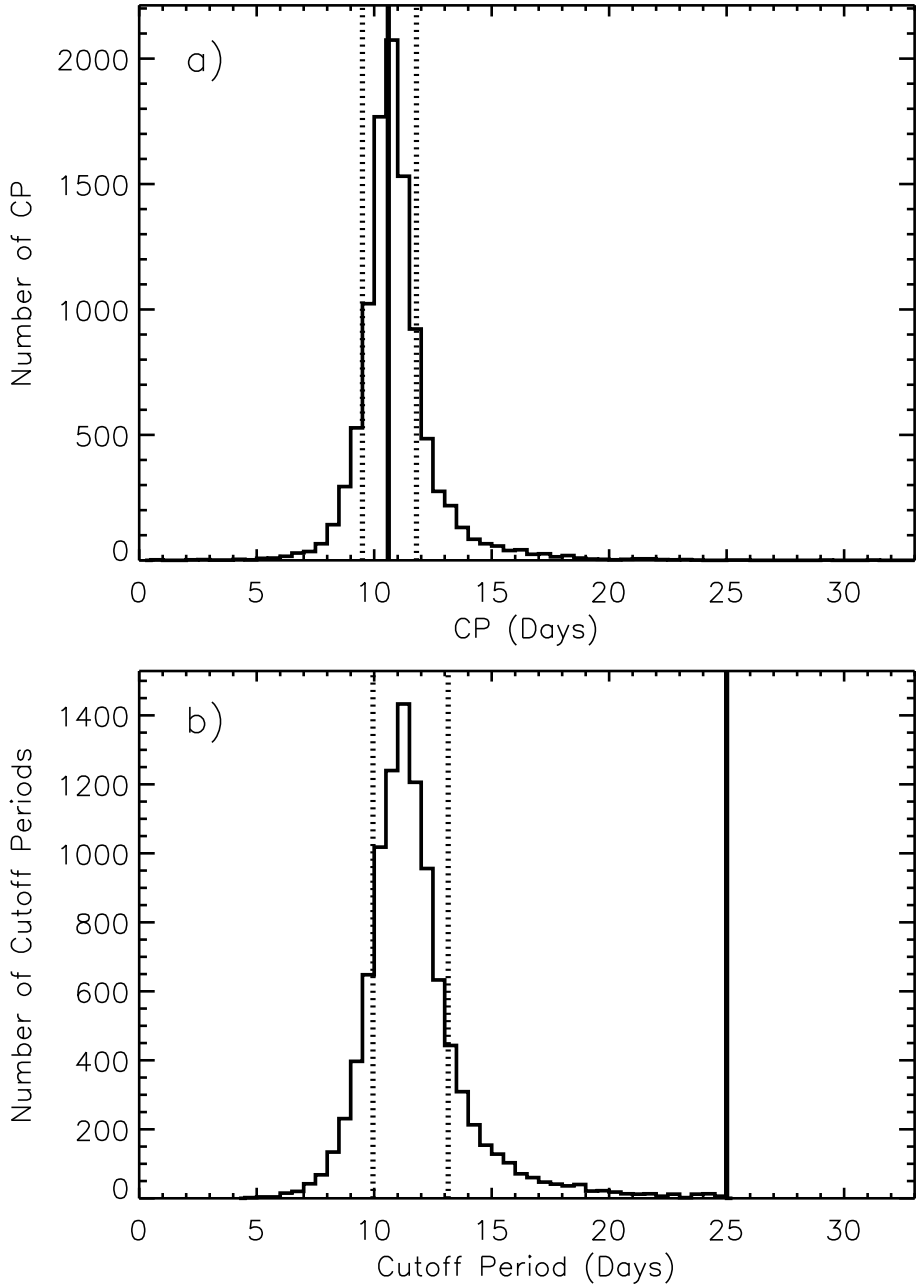}
\caption{{\bf a)} The distribution of circularization periods from
10,000 different binary populations each with 20 binary orbits with
periods between 10 and 50 days. Only initially eccentric binary
orbits ($e_{i} > 0.05$) were considered. The vertical solid line
mark the location of the true circularization period determined from
a ``parent'' population of $\sim$ 20,000 binary orbits (10.6 days).
The interval containing 2/3 of all circularization periods
around the mode is shown by the two vertical dotted lines at 9.6
and 11.8 days. {\bf b)} The distribution of cutoff periods from
the same 10,000 binary populations. The true cutoff period determined
from the ``parent'' population of $\sim$ 20,000 binary orbits is
marked by a vertical solid line at 25.0 days. The vertical dotted
lines at 10.0 and 13.2 days mark the interval containing 2/3 of all
cutoff periods around the mode at 11.3 days. 
\label{hist_nic}}
\end{figure}

Figure~\ref{hist_nic} show the distributions of the observed
circularization periods (\ref{hist_nic}a) and the observed
cutoff periods (\ref{hist_nic}b) derived for the 10,000 different
binary populations {\it excluding} initially circular binary
orbits. The true circularization period is shown in
Figure~\ref{hist_nic}a as a vertical solid line at 10.6 days.
The distribution of circularization periods is slightly asymmetric
with a tail at longer periods. Nonetheless, the true circularization
period falls at the mode of the observed distribution. Thus,
we take an observed circularization period as the best estimate
of the true circularization period for any binary population.

The interval enclosing $2/3$ of the observed circularization
periods on either side of the mode is marked by dotted vertical
lines, and hereafter called the 2/3-interval. We define the
uncertainty on the true circularization period of a binary
population as the $\pm$ period intervals that the true
circularization period can be shifted before the observed
circularization period must be drawn from outside of the
2/3-interval. The errorbars on the true circularization
period can thus be determined by considering two limiting cases:
i) The observed circularization period is located at the
short-period boundary of the 2/3 interval; 1.0 days
shortward of the mode. In that case the observed circularization
period would underestimate the true circularization period
by 1.0 days. We assign a positive errorbar of 1.0 days to
the true circularization period. ii) The observed circularization
period is located at the long period boundary of the 2/3
interval; 1.2 days longward of the mode. In that case the
observed circularization period would overestimate the true
circularization period by 1.2 days. We assign a negative
errorbar of 1.2 days to the true circularization period.

Accordingly, for a binary sample with $\sim$ 20 orbits shorter
than 50 days (e.g. the M35 sample) the uncertainty range around
the circularization period is + 1.0 days and - 1.2 days.

The cutoff period of each binary population is defined here as
the period of the longest period binary with an eccentricity
below 0.05. The distribution of observed cutoff periods in
Figure~\ref{hist_nic}b is also slightly asymmetric with a tail
toward longer periods and a mode of 11.3 days. The true cutoff
period is marked by a vertical solid line at 25 days. Necessarily,
the observed cutoff periods are shorter than the true cutoff
period, and the mode of the distribution falls 13.7 days shortward
thereof.

The measurement uncertainty of the cutoff period is defined
similarly to the circularization period. The 2/3-interval
is marked by dotted vertical lines, and we define the uncertainty
range around our estimate of the true cutoff period as the $\pm$
period intervals that the true cutoff period can be shifted before
the observed cutoff period must be drawn from outside of the
2/3-interval. Accordingly, for a binary sample with 20 orbits
shorter than 50 days (e.g. the M35 sample) the uncertainty range
around the maximum likelihood estimate of the true cutoff period
is + 1.3 days and - 1.9 days.

We wish to emphasize that the maximum likelihood observed
circularization period and the true circularization period
are the same, whereas the maximum likelihood observed cutoff
period is offset from the true cutoff period by 13.7 days.
The latter difference reflects a significant sample size
dependence of the cutoff period that obscures the physical
dependence of the age of the binary population.

In addition the Monte Carlo error analysis gives an error on the
circularization period that is 1.0 day smaller than the error
on the cutoff period. Thus we have shown that the circularization
period is both more accurate and more precise than the cutoff period.

\begin{figure}
\epsscale{0.75}
\plotone{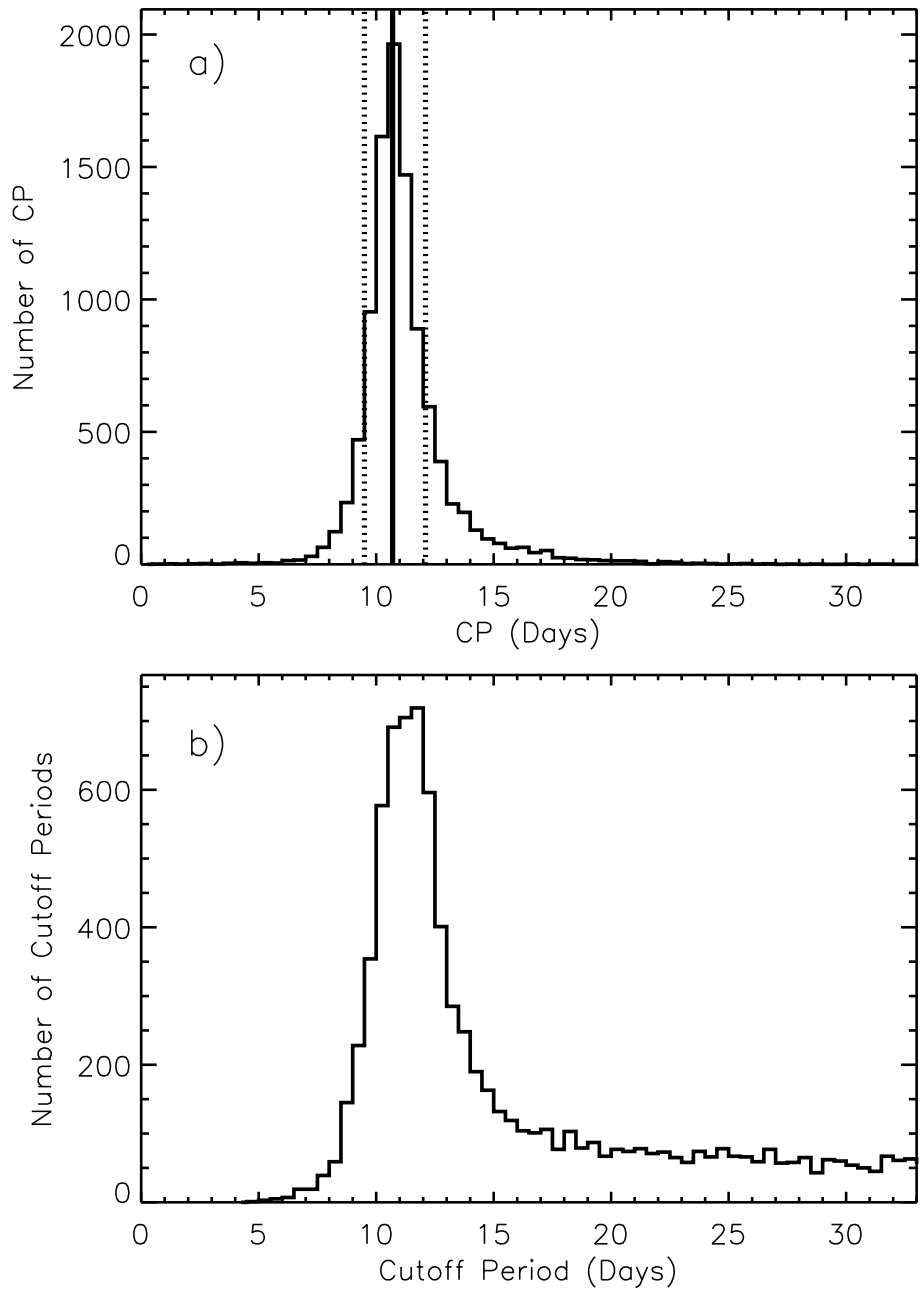}
\caption{{\bf a)} The distribution of circularization periods from
10,000 different binary populations each with 20 binary orbits
with periods between 10 and 50 days. Initially circular binary
orbits ($e_{i} < 0.05$) were included. All vertical lines represent
the same quantities as in Figure~\ref{hist_nic}. The estimated
true circularization period is 10.7 days, and the interval containing
2/3 of all circularization periods around the mode range from
9.7 to 12.2 days. {\bf b)} The distribution of cutoff periods
from the same 10,000 binary populations.
\label{hist_ic}}
\end{figure}

Next we consider binary populations {\it including} initially
circular ($e < 0.05$) binary orbits drawn from the Gaussian
parent eccentricity distribution. Figure~\ref{hist_ic}
shows the distributions of the observed circularization periods
(\ref{hist_ic}a) and the observed cutoff periods (\ref{hist_ic}b)
derived for 10,000 binary populations. All vertical lines
represent the same quantities as in Figure~\ref{hist_nic}.
Note the small effect of initially circular orbits on the
distribution of the observed circularization periods and on
the true circularization period derived from the parent population.
The uncertainty range around the maximum likelihood estimate
of the true circularization period is + 1.0 days and - 1.5 days,
somewhat inflated by the initially circular orbits.

In contrast, the distribution of cutoff periods is drastically
changed by the presence of initially circular binary orbits
($e < 0.05$). The true cutoff period is undefined and a tail
of cutoff periods longward of the distribution mode is produced.

In the presence of binaries with initially low eccentricity
determination of the cutoff period is then dependent on the
judgment of individual observers to define the eccentricity
threshold between circular and eccentric systems and to select
the binary defining the cutoff period. Examples of the effect
of such judgments/selections on the value of the cutoff period
will be given in Section~\ref{7tp} below.

Comparison of Figure~\ref{hist_nic} and Figure~\ref{hist_ic}
clearly demonstrates the robustness of the circularization period
and the vulnerability of the cutoff period. Determination of
the circularization period uses the information from all binary
orbits, circular as well as eccentric, and is thus less vulnerable
to initially circular orbits. The cutoff period is determined
from one binary orbit alone and is thus vulnerable to the
presence of initially circular orbits in the binary population,
contamination from binaries with anomalous evolutionary paths,
and/or binaries that have been falsely classified as cluster
members.

Importantly, the disadvantages of the cutoff period will be
enhanced with larger binary populations, whereas the measurement
uncertainty on the circularization period will decrease with
increasing population size.

To enable ourselves and colleagues to quote errors on
circularization periods to binary populations of different
sizes and ages, we repeated the Monte Carlo error analysis
(allowing for initially circular binary orbits) 9 times.
For each of 3 different sample sizes (10, 20, and 30 binaries)
we tidally evolved each of 10,000 samples so that the modes
of the circularization period distributions were $\sim$ 5,
10, and 15 days. Errors on the circularization periods were
derived in the manner explained above. Table 2 list the derived
errors and can be used as a reference for colleagues studying
tidal circularization in clusters not listed in this paper.

\begin{deluxetable}{cccc}
\tabletypesize{\normalsize}
\tablecaption{Circularization Period Errors
\label{tbl-2}}
\tablewidth{0pt}
\tablehead{
\colhead{Size of} & \colhead{CP = 5 days} &
\colhead{CP = 10 days} & \colhead{CP = 15 days} \\
\colhead{Binary Population} & \colhead{Errors (days)} &
\colhead{Errors (days)} & \colhead{Errors (days)}
}
\startdata
10 Binaries   &   + 1.8  - 1.9   &   + 1.5  - 3.1   &   + 2.3  - 3.2 \\
20 Binaries   &   + 1.2  - 1.2   &   + 1.0  - 1.5   &   + 1.4  - 2.2 \\
30 Binaries   &   + 1.1  - 1.1   &   + 0.8  - 1.1   &   + 1.2  - 1.5 \\
\enddata
\end{deluxetable}

The coefficients $\beta$ and $\gamma$ in the circularization
function (eq. [1]) controls the steepness/slope and the abruptness,
respectively, of the function's transition from zero to non-zero
eccentricities. Values of 0.14 for $\beta$ and 1.0 for $\gamma$
were adopted to minimize the width of the distributions of circularization
periods resulting from the Monte Carlo experiments described above,
and to minimize the sensitivity of the circularization period
to the choice of eccentricity threshold ($e = 0.01$) between
circular and eccentric orbits.


\section{THE CIRCULARIZATION PERIOD IN M35} \label{tpm35}

After motivating, presenting and testing the circularization function
and defining the circularization period, we are ready to apply this
new diagnostic of tidal circularization to the population of binary
orbits in M35.

\begin{figure}
\epsscale{1.0}
\plotone{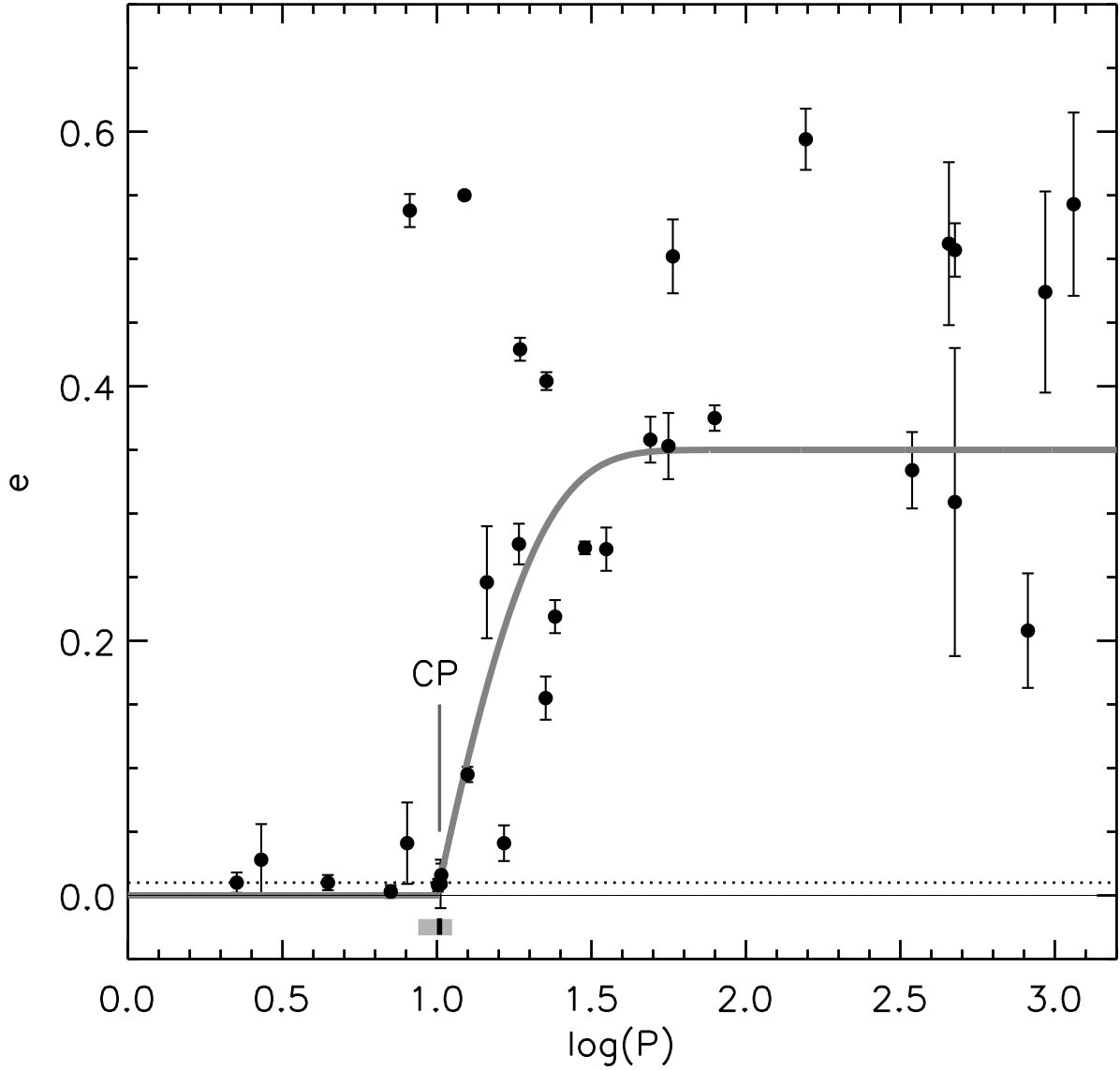}
\caption{The distribution of orbital eccentricity ($e$) as a function
of $log$ orbital period (P) for 32 spectroscopic binary members
of M35. The grey curve represent the best fit circularization function
and the dotted horizontal line mark e = 0.01. The circularization period
(CP) is marked at $\log(P) = 1.0$ (10.2 days) and it's uncertainty
interval overplotted.
\label{m35_log_new}}
\end{figure}

Figure~\ref{m35_log_new} show the period-eccentricity distribution
in the $e-\log(P)$ diagram of M35. The circularization function
resulting in the minimum total absolute deviation between observed
and calculated eccentricities is over-plotted and the locations of
the observed circularization period (CP) and it's uncertainty interval
are marked. The horizontal dotted line marks $e = 0.01$. The observed
circularization period for M35 is $10.2^{+1.0}_{-1.5}$ days. The errors
quoted were derived in Section~\ref{comparing} and are listed in Table 2.

The M35 period-eccentricity distribution show striking similarities
to the artificially generated distributions created using eqs. (3)
and (4) and an initial Gaussian eccetricity distribution. In addition
to the period overlap between eccentric and circular orbits described
in Section~\ref{p-e}, 3 circular binaries with periods of $\sim$ 10
days in M35 nicely represent the most frequent (typical) binaries
located in the high density region at $e \lesssim 0.05$ and $P \sim$
10 days in Figure~\ref{parent_new_cp=10d}. The distribution
of eccentricities of binaries with periods longer than $\sim$ 10
days is consistent with a primordial Gaussian distribution and the
circular binary (4037) with a period 16.49 days is likely to originate
from the low eccentricity wing.

Therefore, M35 is a good example of the advantages of the
circularization period in defining the transition between
eccentric and circular binary orbits. The best fit circularization
function takes advantage of the information provided by all
binary orbits and provide a more robust determination of the
circularization period for the typical binary orbit.


\section{THE CIRCULARIZATION PERIODS OF 7 ADDITIONAL BINARY POPULATIONS} \label{7tp}

\begin{figure}
\epsscale{1.0}
\plotone{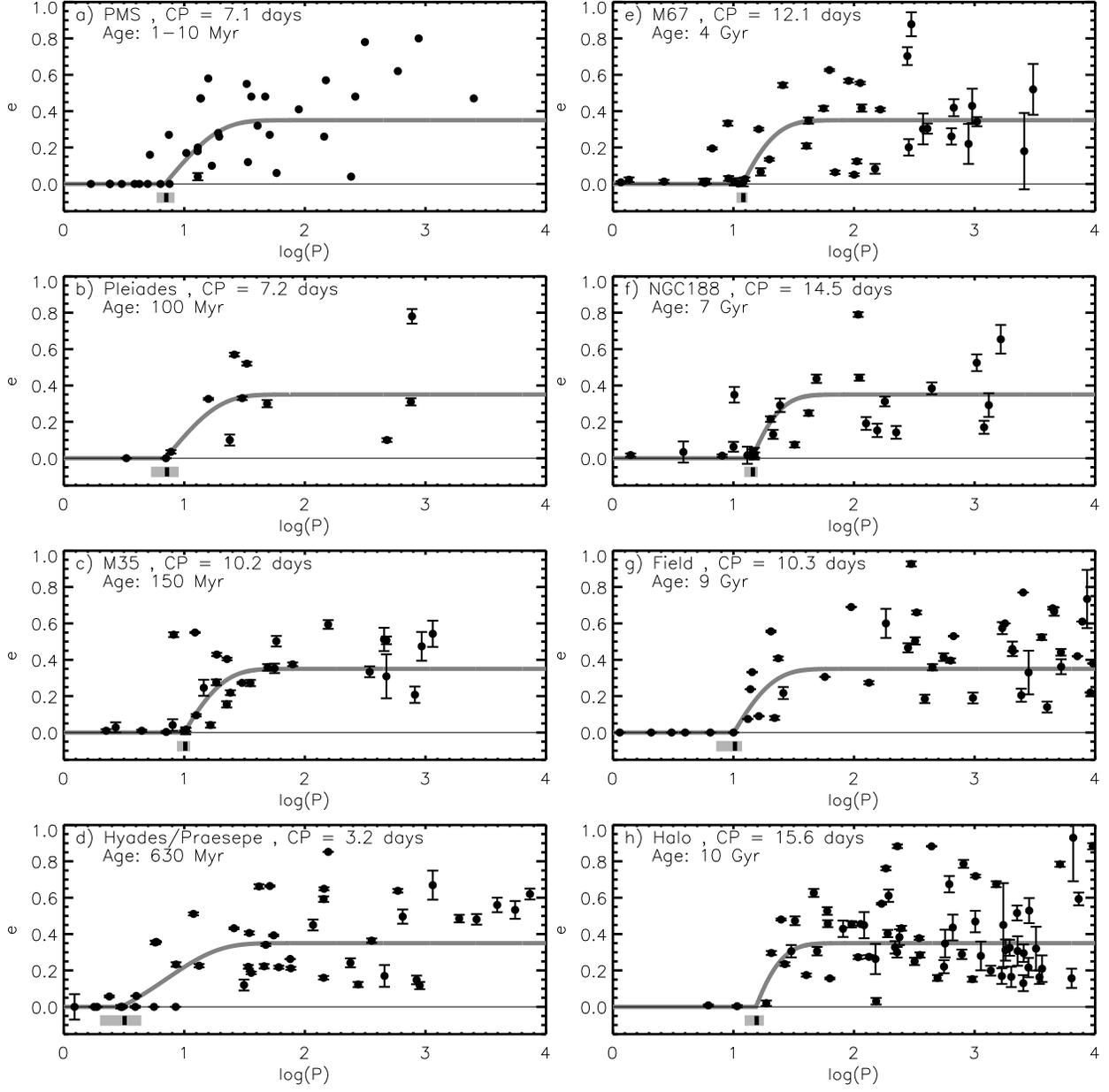}
\caption{The period-eccentricity distributions for 8 late-type
binary populations: {\bf a)} PMS, {\bf b)} Pleiades, {\bf c)}
M35, {\bf d)} Hyades/Praesepe, {\bf e)} M67, {\bf f)} NGC188,
{\bf g)} field, and {\bf h)} halo. Over-plotted each distribution
is the best fit circularization function. A solid horizontal
line mark e(P) = 0.00.
The circularization period (CP) and it's uncertainty are marked
on the period axis by a vertical black line and a grey bar,
respectively. The circularization period and the age of the
binary population are printed in each plot.
\label{all_log}}
\end{figure}

Below we briefly discuss and present the period-eccentricity
distributions and determine circularization periods for 7
additional binary populations spanning ages from $\sim$ 3 Myr
(PMS binaries) to $\sim$ 10 Gyr (Galactic halo binaries).
Figure~\ref{all_log}a-h show the orbital data of each individual
population with the best-fit circularization function over-plotted
and the circularization period with error marked. All results
are listed in Table~\ref{tbl-3}; here we briefly discuss each
population in turn. 

The PMS binary population (Figure~\ref{all_log}a): The sample
of 37 low-mass PMS binaries show the characteristic period
overlap between eccentric and circular orbits. The orbital
parameters are taken from \citet{mca+01}, and references to
individual binaries can be found in their paper. The PMS sample
is not strictly coeval, but cover an age range from $\sim$ 1-10 Myr
\citep{mca+01}. We determine a circularization period of
$7.1^{+1.2}_{-1.2}$ days for the PMS binary sample. This
value should be compared to the 7.56 day cutoff period
determined by Melo et al. Note, that Melo et al. chose to
disregard the circular orbit of binary RX J1301.0-7654a
(P = 12.95 days ($\log(P) = 1.11$), $e = 0.04 \pm 0.02$;
Covino, private communication) based on measurements of
super-synchronous rotation of the individual binary components.
Because the timescale for tidal synchronization is thought to
be shorter than the timescale for circularization \citep{zahn77,hut81},
Melo et al. argue that the circular orbit of RX J1301.0-7654a
might not be a result of tidal circularization. However, as
mentioned by Melo et al., super-synchronous stellar rotation
is an expected result of the transition from the PMS to the
ZAMS due to conservation of angular momentum and a decline
in the efficiency of tidal breaking as the stars contract.
Thus we have chosen to include binary RX J1301.0-7654a.
However, due to the robust nature of our new diagnostic,
excluding binary RX J1301.0-7654a from the PMS sample
causes only a 0.5 day decrease in the circularization period,
whereas the longest period circular orbit would decrease
by 5.39 days (from 12.95 to 7.56 days).

The Pleiades binary population (Figure~\ref{all_log}b):
The eccentricity distribution of 12 binary orbits in the
$\sim$ 100 Myr Pleiades cluster provides poor sampling of
the transition region. Also, unlike any other population
included in this study, the Pleiades features a gap between
the longest period circular and the shortest period eccentric
orbit. By fitting the circularization function we make use of the
information provided by the binary orbits on both sides of
the gap, and determine a circularization period of $7.2^{+1.8}_{-1.9}$
days. This value should be compared to the 7.7 day cutoff
period previously used for the Pleiades cluster. The orbital
parameters shown in Figure~\ref{all_log}b are taken from
\citet{mbm97,mrd+92}. The estimated masses for the binary
primary components falls within $\sim 0.9~M_{\odot} - 1.4~M_{\odot}$.

The M35 binary population (Figure~\ref{all_log}c): The
distribution of orbital eccentricities in M35 (discussed
above) is shown here for the purpose of comparison.

The Hyades/Praesepe binary populations (Figure~\ref{all_log}d):
The sample of 47 binary orbits in the $\sim$ 630 Myr Hyades/Praesepe
twin clusters show 3 eccentric binaries shortward of the longest
period circular orbit. The best fit circularization function is
over-plotted and the circularization period of $3.2^{+1.2}_{-1.2}$
days is marked. The cutoff period for the Hyades/Praesepe sample
has previously been set by the 8.49 day circular binary J331.
J331 consist of two 0.5 $M_{\odot}$ components, and thus is
substantially different from the binaries with approximately
solar-mass components defining the transition regions in other
binary populations. For this reason \citet{mdl+92} questioned
whether binary J331 should be disregarded, as the difference in
mass and stellar structure (depth of convective envelope) might
have significantly altered the timescale of tidal circularization.
We will not comment further on the effect of stellar
mass here, but simply emphasize that while we have included binary
J331, disregarding it leads to no noticeable change in the
Hyades/Praesepe circularization period, compared to a -2.75 days
change from 8.49 to 5.74 days in the Hyades/Praesepe cutoff period.
The orbital parameters shown in Figure~\ref{all_log}d are taken
from \citet{ggg+85,gmg82, gg81,gg78} for Hyades binaries and
\citet{mm99,mrd+92,mwd+90} for Praesepe binaries. The estimated
masses for the the primary components of this sample falls within
$\sim 0.5~M_{\odot} - 1.5~M_{\odot}$.

The M67 binary populations (Figure~\ref{all_log}e):
The sample of spectroscopic binaries in the old open clusters
M67 contain stellar components ranging in evolution from
the unevolved main-sequence to the tip of the giant branch (Mathieu,
priv. comm.). As the rate of tidal circularization depends sensitively
on stellar radius and the depth of the convection zone, we will
consider only binaries in M67 with unevolved primary components.
Figure~\ref{all_log}e shows the distribution of orbital eccentricity
vs. log orbital period for 39 unevolved binaries in M67. Two
eccentric orbits are found shortward of the longest period circular
orbit. The best fit circularization function is over-plotted
and the circularization period of $12.1^{+1.0}_{-1.5}$ days
is marked. The mass-range of the binary primary components
is $\sim 0.9~M_{\odot} - 1.2~M_{\odot}$.

The NGC188 binary populations (Figure~\ref{all_log}f):
As for M67, the binary population in NGC188 contains evolved
as well as unevolved stellar components. Again we will
consider only binaries with unevolved primary components.
The distribution of orbital eccentricity vs. log orbital
period for 27 unevolved binary stars in NGC188 is shown in
Figure~\ref{all_log}f. The binary population contains a
single eccentric system with a period shorter than the
longest period circular binary. The circularization period
of $14.5^{+1.4}_{-2.2}$ days is marked and the best fit
circularization function is over-plotted. The orbital data
shown in Figure~\ref{all_log}f are taken from \citet{mmd04}.
The mass-range of the primary components in this sample is
$\sim 0.95~M_{\odot} - 1.05~M_{\odot}$.

The field binary population (Figure~\ref{all_log}g): The orbital
characteristics of 50 close solar-type binaries from the solar
neighborhood (hereinafter: the field) is shown with the circularization
period of $10.3^{+1.5}_{-3.1}$ days marked and the best fit
circularization function over-plotted. The orbital data are
taken from \citet{dm91}, who adopt an age range of 7-11 Gyr
for the population of field binaries. We adopt 9 Gyr as the
age of the field sample. However, the age of some individual
binary systems are poorly known, thus despite the well-defined
circularization period, the age-ambiguities of this sample makes
it less reliable for probing the evolution of tidal circularization.

The Galactic halo binary population (Figure~\ref{all_log}h):
The orbital characteristics of 61 binaries from the Galactic
halo. The best fit circularization function is over-plotted and
the circularization period of $15.6^{+2.3}_{-3.2}$ days is marked.
The orbital data are taken from \citet{lst+02}. We adopt an age
of 10 Gyr for the halo binary population. From a large sample
of 171 high proper motion spectroscopic binaries, \citet{lst+02}
find no obvious differences between the binary characteristics
in the halo and in the disk populations. The observed frequency
is the same and the period distributions are consistent with the
hypothesis that the binaries are drawn from the same parent
population. Nonetheless, the halo binary population is not strictly
coeval and the stars differ significantly in mass and metallicity
from those of the open cluster/disk populations. Arguably,
comparison of solar-mass/solar-metallicity tidal circularization
theory to the halo cutoff period requires consideration of
these differences.

It is important to note that in the majority of the binary
populations discussed here, circular or marginally circular
binary orbits are observed with periods from tens to hundreds
of days longer than the circularization period. The
selection of unevolved main-sequence binaries for each population
suggest that these long period ($P \sim 10^1 - 10^2$ days)
low eccentricity ($e \sim 0.05$) orbits are not a result of
accelerated tidal circularization due to evolved primary
components. The existence of long-period low eccentricity
binaries in the PMS, Pleiades and M35 samples is a strong
indication that binary stars can form with circular or
marginally circular orbits, consistent with an initial
Gaussian distribution of orbital eccentricities.

The binary populations shown in Figure~\ref{all_log} also
show a high frequency of high-eccentricity orbits at periods
shorter than the one or more circular orbits. The ability
to reproduce such systems from binaries in the high-eccentricity
tail of the initial Gaussian distribution by numerically
integrating current theoretical models of tidal circularization
\citep[Section~\ref{comparing} this paper]{dmm92}, provide
further support for the existence of an initial Gaussian
distribution of orbital eccentricities.

\begin{deluxetable}{cccc}
\tabletypesize{\normalsize}
\tablecaption{Distribution of Circularization Periods with Population Age
\label{tbl-3}}
\tablewidth{0pt}
\tablehead{
\colhead{Binary Population} & \colhead{$\log(Age)$} &
\colhead{Circularization Period} \\
\colhead{} & \colhead{(Gyr)} & \colhead{(days)}
}
\startdata
PMS Binaries    & -2.5 &  $7.1^{+1.2}_{-1.2}$ \\
Pleiades        & -1.0 &  $7.2^{+1.8}_{-1.9}$ \\
M35             & -0.8 & $10.2^{+1.0}_{-1.5}$ \\
Hyades/Praesepe & -0.2 &  $3.2^{+1.2}_{-1.2}$ \\
M67             &  0.6 & $12.1^{+1.0}_{-1.5}$ \\
NGC188          &  0.8 & $14.5^{+1.4}_{-2.2}$ \\
Field Binaries  & 0.95 & $10.3^{+1.5}_{-3.1}$ \\
Halo Binaries   & 1.00 & $15.6^{+2.3}_{-3.2}$ \\
\enddata
\end{deluxetable}


\section{EVOLUTION OF TIDAL CIRCULARIZATION} \label{evol}

While different tidal circularization theories agree
that the timescale of circularization depends strongly
on stellar separation, and that consequently a transition
from eccentric to circular orbits is expected at a well-defined
binary period, the different theories describe different
mechanisms for the tidal dissipation leading to circularization.
The mechanism and efficiency of tidal dissipation in a given
model reflects itself in the model's prediction of the rate
of circularization and the evolution of the tidal circularization
period. Observational determinations of the circularization
periods in binary populations of different ages provide a
critical test of the dissipation mechanism(s) responsible.

Detailed discussion of the distribution of circularization
cutoff periods with age, and its ability to constrain tidal
circularization theory, was given by \citet{ws02},\citet{mca+01},
and \citet{mdl+92}. However, the addition of the new large
binary populations of M35 and NGC188 \citep{mmd04}, the
introduction of a new robust measure of tidal circularization
(the circularization period) with measured uncertainties,
and the recent theoretical work on dynamical tides in late-type
stars \citep[e.g.][]{ws02}, motivate a fresh look at how the
theory compares to the observations. 

\begin{figure}
\epsscale{1.0}
\plotone{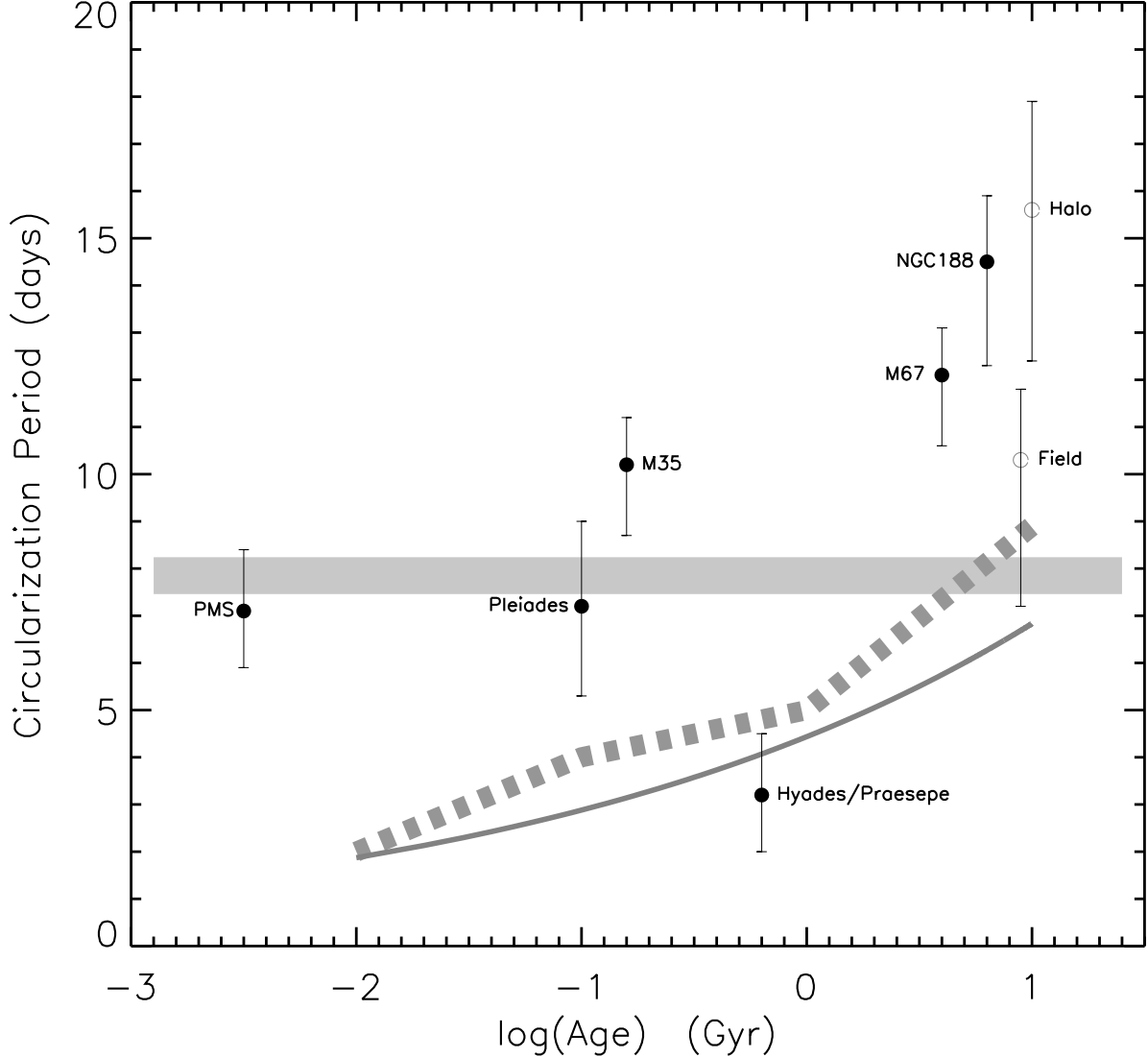}
\caption{The distribution of circularization periods with age
for 8 late-type binary population (solid/open circles). Errorbars
represent the uncertainties on the circularization periods derived
in Section~\ref{comparing}. The solid curve shows the predicted
cutoff period as a function of time based on main-sequence
tidal circularization using the revised equilibrium tide theory
by \citet{zahn89} \citep{cc97}. The broad dashed band represent
the predicted cutoff period period for initially super-synchronous
$1~M_{\odot}$ stars calculated in the framework of the dynamical
tide model including resonance locking \citep{ws02}. The horizontal
grey band represents the prediction by \citet{zb89} in which tidal
circularization is significant only during the PMS phase.
\label{logAge_cp}}
\end{figure}

We show in Figure~\ref{logAge_cp} the circularization periods
for all binary populations discussed in this paper (solid/open
circles) at their respective ages. Errorbars show the uncertainties
on the circularization periods based on the Monte Carlo experiments
described in Section~\ref{comparing}. 

With the exception of the Hyades/Praesepe population, the
circularization periods of the coeval solar-type binary
populations show a steady increase with age from the
PMS and early main-sequence to the late main-sequence
phase. The circularization period of the halo and field
binaries follow this trend. However, due to the non-coevality
and uncertainty in age of the field sample, and the differences
in mass and metallicity of the stars in the halo sample,
we choose not to include their circularization periods
in the subsequent analysis.

The small value of the circularization period of the
Hyades/Praesepe sample can be explained in part by the
high number of short period eccentric binaries, in particular
the two highly eccentric short period binaries (Hyades:
HD30738, $P=5.75$ days, $e=0.354$; Praesepe: KW181, $P=5.87$
days, $e=0.357$). However, even in the absence of these
two systems the Hyades/Praesepe circularization period
will be 5.9 days and thus still deviate from the overall
trend of increasing circularization periods with increasing
age. We note that \citet{dm91} draw attention to an excess
of short period binaries ($\log(P) < $ 1 day) in the Hyades
period distribution when compared to distribution from
their G-dwarf sample.

The predictions of main-sequence tidal circularization
in the framework of both the equilibrium tide theory and the
dynamical tide theory are also shown in Figure~\ref{logAge_cp},
as is the prediction by \citet{zb89} including PMS tidal
circularization. Note that these different theoretical
predictions displayed in Figure~\ref{logAge_cp} use different
criteria for determination of the cutoff period. \citet[equilibrium
tide theory]{cc97} define the cutoff period at a given age
as the longest orbital period for which the relative variation
in the eccentricity is 0.5\% of the initial value ($e_{ini}
= 0.30$). \citet[dynamical tide theory]{ws02} define the cutoff
period as the longest orbital period for which a binary with
$1~M_{\odot}$ components has been circularized to $e < 0.01$.
As such the predictions of Witte \& Sovonije are upper limits
on the circularization period. \citet[PMS tidal theory]{zb89}
define the cutoff period as the period at which a binary with
component masses between $0.5~M_{\odot} - 1.25~M_{\odot}$ and
initial orbital eccentricity of 0.2 or 0.3 circularizes to $e
< 0.005$. Recall that the circularization periods displayed in
Figure~\ref{logAge_cp} represent the orbital period at which a
binary orbit with initial eccentricities in the range 0.2 -
0.5 evolves to $e = 0.01$ at the age of the population.

{\it The equilibrium tide theory}:
Let us consider first the equilibrium tide theory using the
dissipation mechanism by \citet{zahn89}. The equilibrium tide
theory predicts active tidal circularization throughout the
main-sequence phase, leading to longer cutoff periods with
increasing population age (solid curve in Figure~\ref{logAge_cp}).
However, as pointed out by \citet{cc97}, an artificial
enhancement of the turbulent dissipation is necessary
to fit the observed cutoff periods. Comparison with the new
circularization periods, again with the exception of the
circularization period of the Hyades/Praesepe population,
show a similar discrepancy between the observations and the
theoretical predictions of pure main-sequence tidal circularization.

The grey horizontal band in Figure~\ref{logAge_cp} shows the
prediction by \citet{zb89} of tidal circularization including
the PMS phase. Based on the reduced rate of tidal circularization
derived by \citet{zahn89}, Zahn \& Bouchet suggested that all
tidal circularization occurs during the PMS phase when the stars
have larger radii and deeper convective envelopes. They found
that for main-sequence stars the efficiency of turbulent
dissipation derived from the equilibrium tide theory is so
small that tidal circularization following the PMS phase is
negligible despite the Gyr timescales available. Their PMS tidal
circularization theory predicts a range of cutoff periods between
$\sim$ 7.2 - 8.5 days for binaries with components with masses
between 0.5 - 1.25 $M_{\odot}$ and initial eccentricities of either
0.2 or 0.3. Accordingly, all late-type main-sequence binaries
should have circular orbits for period less than $\sim$ 8 days
and (primordial) eccentric orbits at longer periods, independent
of age.

The hypothesis that PMS tidal circularization alone sets a tidal
cutoff period, independent of age, is $not$ consistent with the
observed distribution of circularization periods. We performed a
$\chi^2$ test fitting different constant theoretical circularization
periods to the distribution of observed circularization periods.
We find that there is less than a 1\% probability that the observed
circularization periods (excluding or including the field and
halo samples) derive from a model predicting no evolution of the
circularization periods with age. Repeating the $\chi^2$ test
for all binary populations but the Hyades/Praesepe population
also tells us that there is less than 1\% chance that the PMS,
Pleiades, M35, M67, and NGC188 populations derive from a model
predicting pure PMS circularization.

{\it The dynamical tide theory}: The predicted evolution
of main-sequence tidal circularization using the dynamical
tide theory with inclusion of resonance locking \citep{ws02}
is shown in Figure~\ref{logAge_cp} as a broad dashed band.
The width of the dashed band represents the range of predicted
evolutions of the cutoff period for binaries with stellar components
rotating at super-synchronous ($\Omega = 2 \omega_{p}$), pseudo-synchronous
($\Omega = \omega_{p}$), and orbital ($\Omega = \omega$) speeds.
$\Omega$, $\omega_{p}$, and $\omega$ denote the initial stellar
angular velocity, the orbital angular velocity at periastron, and
the average orbital angular velocity, respectively. The masses of
the binary components used by Witte \& Savonije are 1 $M_{\odot}$.

Regardless of stellar rotation velocity, the dynamical tide theory
with inclusion of the resonance locking mechanism \citep{ws02}
appears to predicts slightly more efficient tidal circularization
than the equilibrium tide theory throughout the main-sequence phase.
However, \citet{ws02} define the cutoff period as the longest
orbital period for which a binary has been circularized to $e
< 0.01$. This definition implies that the predicted cutoff periods
are likely set by binaries with low initial orbital eccentricities,
and thus are not directly comparable to the predictions by \citet{cc97}.
Still there is a discrepancy between the predicted level of circularization
and the distribution of observed circularization periods. For the
special case of very slowly rotating stars, the calculations of
\citet[not shown in Figure~\ref{logAge_cp}]{ws02} offer a near
match to the circularization periods of Pleiades, M35, M67 and
NGC188. However, the required stellar rotation periods are of
order 100 days for stars in close binary systems with ages between
0.1 and 10 Gyr. Such very slow rotations would be unexpected and
thus needs to be observationally established.

The slopes of the main-sequence tidal circularization models
match the increase in circularization period seen in
Figure~\ref{logAge_cp}. Thus if the efficiency of tidal dissipation
were to be artificially enhancement for both the equilibrium
tide and the dynamical tide theories, the predicted main-sequence
evolution for both theories would be in agreement with the
observed increase in circularization period for all populations
but the Hyades/Praesepe. 

{\it The hybrid scenario}:
The lack of success of both PMS and main-sequence
tidal circularization theories to account for the observations
of tidal circularization motivated the suggestion of a
``hybrid scenario'' \citep{mdl+92}. The hybrid scenario is
heuristic and was motivated by a distribution of cutoff
periods that appeared age-independent for populations younger
than $\sim$ 1 Gyr while showing a positive correlation with
age for populations older than $\sim$ 1 Gyr. The hybrid scenario
suggest that tidal circularization in binary populations younger
than $\sim$ 1 Gyr derive from PMS tidal circularization, and that
after the passage of $\sim$ 1 Gyr the integrated main-sequence
tidal circularization begins to circularize binaries with orbital
periods of $\sim$ 10 days.

Considering the new distribution of circularization periods
the need for a hybrid scenario is uncertain. With the exception
of the Hyades/Praesepe population the observed circularization
periods increase with age from the PMS and throughout the
main-sequence phase. A $\chi^2$ test shows that there is less
than 1\% probability that the circularization periods for
populations younger than $\sim$ 1 Gyr derive from a model
predicting no main-sequence tidal circularization. Excluding
the Hyades/Praesepe sample result in $\chi^2$ value of $\sim$ 1.
Thus, the current distribution of circularization periods for
the youngest binary populations cannot distinguish between
a model predicting an age-independent circularization period
or a model predicting a continuous increase in the circularization
period from the PMS onward.

If tidal circularization is in fact active throughout the
main-sequence phase, the question remains which if any of the
suggested dissipation mechanisms is responsible. Either
the equilibrium or the dynamical tide mechanisms, or perhaps
both in combination, could be responsible for the circularization
of binary orbits during the main-sequence phase. However,
unless both mechanisms are at work simultaneously, more
efficient tidal dissipation is needed in both the equilibrium
and the dynamical tide theories. We note that PMS tidal
circularization is still needed to explain the 7.1 day
circularization period of the PMS sample, and that an
increase in the efficiency of turbulent dissipation of
the equilibrium tide theory presumably will affect
circularization during the PMS as well.

Alternatively, assuming pure PMS tidal circularization, the
differences in circularization periods among the binary populations
in Figure~\ref{logAge_cp} could be due to differences in initial
stellar and circumstellar conditions. \citet{gd98} speculate
that perhaps the differences in mass and/or metallicity between
the younger disk binaries and the older disk and halo binaries
could be the cause of differences in cutoff periods set during
the PMS phase. While the ranges of primary stellar masses and
metallicities of the populations included in this study are similar,
other stellar/circumstellar and binary/circumbinary parameters
might be of importance to the efficiency and duration of tidal
circularization during the PMS phase. The strong dependence
of the efficiency of circularization on stellar rotation displayed
in the dynamical tide theory by \citet{ws02} is a powerful
example of how differences in initial conditions may affect
the evolution of tidal circularization.

It is difficult to draw secure conclusions about the evolution
of tidal circularization based on only 6 circularization
periods. The future observational goal is to further populate
the age vs. circularization period diagram (Figure~\ref{logAge_cp})
with reliable circularization periods at carefully selected
ages based on large samples of binary orbits from homogeneous
coeval binary populations. Following the present study and
the recent publication of binary orbits in NGC188 \citep{mmd04},
we plan to contribute (work in progress) such circularization
periods at the ages of the young open cluster M34 ($\sim$ 250 Myr)
and the intermediate age open cluster NGC6819 ($\sim$ 2.5 Gyr).
By inspection of Figure~\ref{logAge_cp} it is evident that these
future circularization periods will greatly improve the observational
constraint on future theoretical models and predictions
of tidal circularization in close late-type binary stars.

The future goal for theory of tidal circularization in
late-type main-sequence stars is to construct models that
combine PMS tidal circularization with main-sequence tidal
circularization, and predict more efficient tidal circularization
in binaries with main-sequence components to account for
the observed evolution of circularization.


\section{SUMMARY AND CONCLUSIONS} \label{conclusions}

In a coeval population of binary stars the closest binaries
tend to have circular orbits, while wider binaries have orbits
with a distribution of non-zero eccentricities. Determination
of the orbital period to which binaries have been circularized
is an important observational constraint on theoretical models
seeking to explain the dissipation mechanism(s) responsible for
tidal circularization. Coeval populations of short-period
binary orbits are therefore key observational contributions.

However, despite homogeneity in age, mass and metallicity,
a population of short-period binaries may be small in size
and will consist of systems with a distribution of initial
eccentricities. The size of a binary population will affect
the uncertainty on its circularization period. The initial
eccentricity of a circular binary with a period equal to the
circularization period is unknown, but can be estimated
through the use of theoretical models. Proper comparison
between observed and theoretical circularization periods thus
require measurement uncertainties on the observed circularization
periods and information about the initial eccentricities
of the binaries that circularize at those periods.

The primary contributions of this study are: 1) a new population
of solar-type spectroscopic binary orbits for the 150 Myr
open cluster M35; 2) a new diagnostic to determine the period
of circularization of the most frequently occurring binary;
3) an evaluation of the performance of this new diagnostic
providing measurement uncertainties on the resulting circularization
periods and information about the initial eccentricity of the
binaries that circularize at that period; and 4) a comparison
of the distribution of circularization periods from 8 binary
populations to the predictions from theoretical models.

The new sample of 32 spectroscopic binary orbits
in the ``Pleiades age'' ($\sim$ 150 Myr) open cluster M35
greatly improves the constraint on tidal circularization at
the early stage of the main-sequence phase. Prior to this
study, the small sample of binary orbits in the Pleiades
provided only very rough limits on the transition between
eccentric and circular orbits at this age.

Monte Carlo simulations of tidal circularization using
the equilibrium tide theory by \citet{zahn77} were used
to create artificial $e-\log(P)$ distributions. These
simulations, together with the $e-\log(P)$ distributions
of M35 and 7 additional published binary populations, were
used to critically assess the adequacy of the tidal
circularization cutoff period (the period of the longest
period circular binary) to measure the degree of tidal
circularization. We conclude that the cutoff period is not
an optimal measure of the transition between eccentric and
circular orbits because of the following disadvantages:

\begin{itemize}

\item The simulations of tidal circularization using the
equilibrium tide theory by \citet{zahn77} show that the
cutoff period is likely to originate from the low eccentricity
tail of an initial Gaussian eccentricity distribution. Therefore,
the cutoff period does not represent the period of circularization
of the most frequently occurring binary orbit.

\item The expected value of the observed cutoff period of
a population of binaries varies with the size of the population.

\item The cutoff period is vulnerable to the presence of
circular orbits that are not a measure of tidal effects, such
as primordial circular binary orbits or orbits of binaries
with anomalous evolutionary paths.

\item Determination of the cutoff period makes no use of
the information provided by eccentric orbits in the $e-\log(P)$
diagram.

\end{itemize}

Motivated by the need for a more robust determination of
the orbital period at which a binary orbit with the most
frequent initial orbital eccentricity circularizes, we introduce
the circularization function and the circularization period.
The functional shape of the circularization function (eq. [1])
is fixed and mimics the transition from eccentric to circular
orbits in the $e-\log(P)$ diagram (Figure~\ref{parent_new_cp=10d}).
The circularization period is defined as the period for which
the circularization function has a value of 0.01 ($e(CP) = 0.01$).
The location in period of the circularization function thus determines
the circularization period. The advantages of the circularization
function/period are:

\begin{itemize}

\item Numerical simulations of tidal circularization using the
equilibrium tide theory by \citet{zahn77} show that the circularization
period is the period at which a binary with the most frequent
initial eccentricity ($e \sim 0.35$) evolves to $e = 0.01$ at the
age of the population.

\item The circularization period is determined from all binary orbits
(circular and eccentric), and so is less vulnerable to primordial
circular binaries or binaries with anomalous evolutionary paths.

\item The observed circularization period is the best estimate
of the true circularization period independent of the size of
the binary population.

\end{itemize}

Monte Carlo error analysis allows us to examine the distribution
of the circularization periods. From this analysis we estimate
and provide the uncertainty on the circularization period as a
function of the size of the binary sample and of the value of the
circularization period. We note that even in the absence of initial
circular binaries and/or contamination from binaries with anomalous
evolutionary paths, the Monte Carlo error analysis show that the
uncertainty on the circularization period is smaller than the uncertainty
on the cutoff period. Thus we conclude that the circularization
period is both more accurate and more precise than the cutoff
period.

We present the circularization periods for M35 and 7 additional
binary populations ranging in age from $\sim$ 3 Myr to $\sim$
10 Gyr. With the exception of the circularization period of the
Hyades/Praesepe population, the distribution of circularization
periods with age show a steady increase from the PMS to the late
main-sequence.

We compare the distribution of circularization periods to the
predictions by theoretical models. The models of main-sequence
tidal circularization using either the equilibrium tide theory
\citep{zahn89} or the dynamical tide theory with resonance locking
\citep{ws02} both predict longer cutoff periods with increasing
population age in agreement with the trend in the distribution
of circularization periods. {\bf However, the predicted circularization
periods fall significantly below the observed circularization
periods.} This suggest that the efficiency of the dissipation
in these models is too low.

A model including PMS tidal circularization \citep{zb89}
predicts negligible circularization during the main-sequence
phase. This prediction is not consistent with the observed
distribution of circularization periods. 

The need for a hybrid scenario as proposed by \citep{mdl+92}
is uncertain. The current distribution of circularization
periods for the youngest binary populations cannot distinguish
between a model predicting an age-independent circularization
period or a model predicting a continuous increase in the
circularization period from the PMS onward.

The goal for future theories of tidal circularization in
late-type main-sequence binaries is to combine PMS and
main-sequence circularization in one model that can account
for the circularization periods observed at all ages.


\acknowledgments

We are very grateful to the referee for thoroughly reading the
paper and making suggestions for important improvements. We are
thankful to the University of Wisconsin - Madison Astronomy Department,
to NOAO, and to the WIYN Telescope Director George Jacoby for
the time granted on the WIYN telescope. We would like to express
our appreciation for exceptional and friendly support of the WIYN
Observatory staff, and to Christopher J. Dolan for designing the
datareduction pipeline used to measure stellar radial velocities
and for obtaining the first epochs in the dataset presented.
We thank Jeff Percival and Charles Corson for developing the
infrastructure to make remote observing with the WIYN telescope
easy and efficient, Ellen Zweibel for granting us liberal access
to the Ariadne supercomputer at UW Madison, and Imants Platais for
help with identification of m35 stars in the 2MASS survey. We
thank Joel Robbins (UW Mathematics Dept.) and Keivan Stassun for
help in the search for the circularization function. This work has
been supported by NSF grant AST 97-31302 and by a Ph.D fellowship
from the Danish Research Agency (Forskningstyrelsen) to S.M. 

\newpage




\end{document}